\documentclass[ aip, jcp, amsmath, amssymb, reprint]{revtex4-1}
\usepackage{graphicx}
\usepackage{dcolumn}
\usepackage{bm}
\usepackage{docs}
\usepackage{bm}
\usepackage{textcomp}
\usepackage[colorlinks=true,linkcolor=blue, citecolor=blue, urlcolor=blue]{hyperref}
\usepackage[latin1]{inputenc}
\usepackage[dvipsnames]{xcolor}
\usepackage{todonotes}

\newcommand{\fhi}{ \affiliation{Fritz-Haber-Institut der Max-Planck-Gesellschaft, Faradayweg 4-6, D-14195, Berlin, Germany} }
\newcommand{\tum}{ \affiliation{Department Chemie, Technische Universit\"at M\"unchen, Lichtenbergstr. 4, D-85748,  Garching, Germany} }
\newcommand{\yle}{ \affiliation{Department of Chemistry, Yale University, New Haven, CT 06520, United States} }
\newcommand{\nan}{ \affiliation{Nano Structural Materials Center, School of Materials Science and Engineering, Nanjing University of Science and Technology, Nanjing 210094, Jiangsu, China} }
\newcommand{\lux}{ \affiliation{Physics and Materials Science Research Unit, University of Luxembourg, L-1511 Luxembourg }}

\begin{document}
\title{Adsorption structures and energetics of molecules on metal surfaces: Bridging experiment and theory}

\author{Reinhard J. Maurer} \yle\tum
\author{Victor G. Ruiz} \fhi
\author{Javier Camarillo-Cisneros} \fhi
\author{Wei Liu} \nan
\author{Nicola Ferri} \fhi
\author{Karsten Reuter} \tum
\author{Alexandre Tkatchenko} \fhi\lux
\email{tkatchenko@fhi-berlin.mpg.de}

\begin{abstract}
Adsorption geometry and stability of organic molecules on surfaces are key parameters that determine the observable properties and functions of hybrid inorganic/organic systems (HIOSs). Despite many recent advances in precise experimental characterization and improvements in first-principles electronic structure methods, reliable databases of structures and energetics for large adsorbed molecules are largely amiss. In this review, we present such a database for a range of molecules adsorbed on metal single-crystal surfaces. The systems we analyze include noble-gas atoms, conjugated aromatic molecules, carbon nanostructures, and heteroaromatic compounds adsorbed on five different metal surfaces. The overall objective is to establish a diverse benchmark dataset that enables an assessment of current and future electronic structure methods, and motivates further experimental studies that provide ever more reliable data. Specifically, the benchmark structures and energetics from experiment are here compared with the recently developed van der Waals (vdW) inclusive density-functional theory (DFT) method, DFT+vdW$^{\mathrm{surf}}$. In comparison to 23 adsorption heights and 17 adsorption energies from experiment we find a mean average deviation of 0.06~\AA{} and 0.16~eV, respectively. This confirms the DFT+vdW$^{\mathrm{surf}}$ method as an accurate and efficient approach to treat HIOSs. A detailed discussion identifies remaining challenges to be addressed in future development of electronic structure methods, for which the here presented benchmark database may serve as an important reference.
\end{abstract}

\maketitle

\tableofcontents
\makeatletter
\let\toc@pre\relax
\let\toc@post\relax
\makeatother 

\newpage

% 
% \listoftodos
% \newpage
\section{Introduction}

The interaction of organic materials and molecules with metal surfaces is of widespread interest to both fundamental science and technology. The eventual control of the functionality of the formed hybrid inorganic-organic systems (HIOSs) has potential applications to a variety of fields ranging from functionalized surfaces, to organic solar cells~\cite{Hoppe2011}, molecular electronics~\cite{Mirkin1992}, nanotechnology~\cite{Joachim2000, Russew2010}, and medical implantology~\cite{Bauer2013}. A bottom-up approach of molecular nanotechnology promises a potential route to overcome size limitations of nanoscale devices constructed with traditional top-down approaches such as lithography based device design. An important prerequisite to such an approach is the ability to control and manipulate the structure and interactions of individual molecular building blocks mounted on well-defined surfaces. This can only be achieved with the expertise to fully characterize the adsorption geometry and fully understand the, sometimes subtle, interplay of interactions that lead to a particular molecule-substrate binding strength. 

This fundamental aspect has seen a rapid development over the last few years in the surface science context, \textit{i.e.} for the adsorption of large and complex organic adsorbates at close to ideal single-crystal surfaces and under the well-defined conditions of ultrahigh vacuum and low temperature~\cite{Rosei2003, Barlow_Raval_2003, Tautz2007, Barth2007}.
On the one hand, this development has been driven by significant methodological advancements in individual experimental techniques, such as improvements in the resolution of Scanning Tunneling Microscopy-based (STM) imaging techniques~\cite{Hapala2014} or surface-enhanced Raman spectroscopy~\cite{Nie1997}. Many of these approaches are used complementary to each other or are being combined to enhance resolution, such as is the case for STM tip-enhanced local Raman experiments~\cite{Pettinger2012,Noguez2015}. On the other hand, theoretical developments have been driven by advancements in predictive-quality first-principles calculations. The most important of which in the context of HIOSs is the efficient incorporation of long-range dispersion interactions into semi-local or hybrid Density-Functional Theory (DFT) calculations~\cite{Tkatchenko2010, Liu2014}. 

When applied to HIOSs, first-principles calculations face a multitude of challenges, even for single-molecules with simple adsorbate geometry. Localized molecular states of the adsorbate and the delocalized metal band structure have to be described on equal footing, whereas most currently used (and computationally tractable) approximations to the exchange-correlation (xc) functional in DFT are optimized to perform well for either one or the other. This is aggravated by the need to additionally describe dispersive interactions, which are generally not contained in such lower-rung functionals, but can easily play a dominant role \textit{e.g.} in the adsorption of conjugated or aromatic molecules. The resulting interplay of dispersion interactions, wave-function hybridization, Pauli repulsion, and charge-transfer at such interfaces demands an efficient and accurate electronic structure description that is able to provide a well-balanced account of this wide range of interactions.

\begin{figure}
\includegraphics[width=\columnwidth]{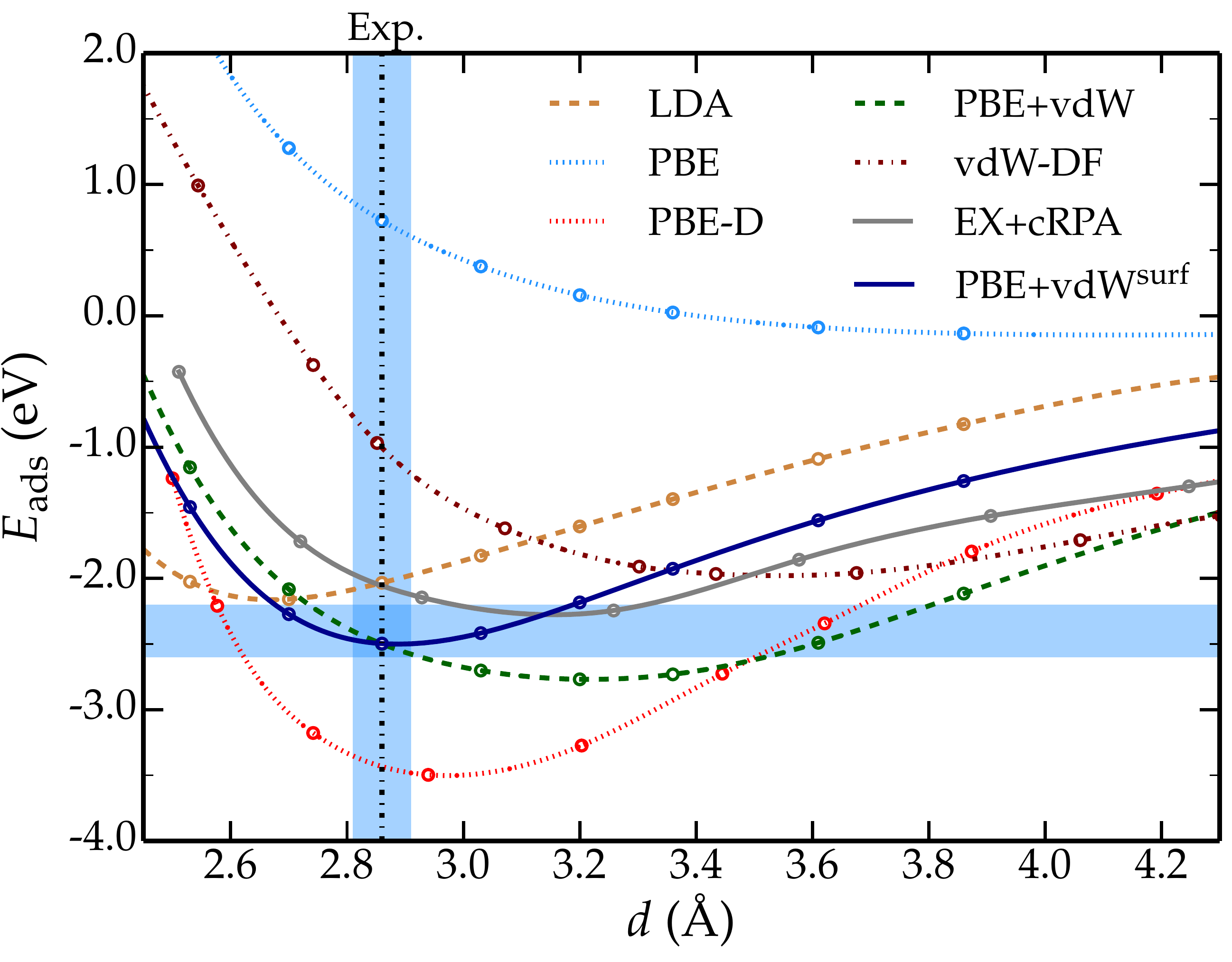}
\caption{\label{fig:intro} Adsorption energy curve of PTCDA adsorbed at Ag(111). Shown are the results of different DFT methods and dispersion correction approaches. Experimental results from X-ray standing wave measurements~\cite{Hauschild2010} and estimated from TPD data of the smaller analogue molecule NTCDA~\cite{Tkatchenko2010,Ruiz2012} are shown as blue bars. A detailed analysis of this figure can be found in section \ref{chapter-interpretation}. Based on Fig. 1 of Ref. \onlinecite{Ruiz2012}.}
\end{figure}

A number of strong contenders to this goal have been suggested, such as the recently developed DFT+vdW$^{\mathrm{surf}}$ method~\cite{Ruiz2012,Liu2015}. However, there is still room for further improvement. This situation is nicely illustrated in Fig.~\ref{fig:intro}, which compiles the predictions of the adsorption height and adsorption energy for 3,4,9,10-perylene-tetracarboxylic acid (PTCDA) adsorbed on Ag(111) as obtained by a variety of pure and dispersion-corrected xc-functionals ranging from local xc-approximations to non-local correlation based on the Random Phase Approximation (RPA)~\cite{Rohlfing2008}.
Quite symptomatic for HIOSs in general, a wide spread in the predicted quantities is obtained. Without an independent experimental reference, we would thus not be able to unambiguously identify the merits and deficiencies of the individual approximations and models that underlie the methods in Fig.~\ref{fig:intro}.

Reliable experimental data on geometric structure and energetics of organic-inorganic interfaces are thus urgently needed as a reference for validation and benchmarking of new and improved electronic structure methods. This reference data must thereby properly match what is calculated. Many experimental techniques to characterize the structure and energetics of HIOSs are based on probing the statistics of an ensemble of adsorbates. It is the very complex mixture of adsorbate-surface and lateral adsorbate-adsorbate interactions that drives the aspired self-assembly of HIOSs that also gives rise to a rich structural phase behavior as a function of temperature and coverage~\cite{Gahl2013, Diller2014,Mueller2016, Willenbockel2015}. Care has to be taken to compare consistent arrangements at the surface in experiment and theory. The targeted molecules can furthermore exhibit a pronounced element of flexibility, disorder, and structural anharmonicity~\cite{Ditze2014, Marbach2014, Maurer2016}. This gives rise to significant finite-temperature effects that also need to be carefully disentangled~\cite{Mercurio2013, Maurer2016}.

A correct interpretation of experiments is therefore of vital importance. If accomplished, the formulation of well-balanced, diverse sets of benchmark systems can then facilitate methodological improvements as has been shown by the success of the S22 dataset for intermolecular interactions of gas phase molecules~\cite{Jurecka2006}, or the C21/X16/X23 databases of molecular crystals~\cite{Otero-de-la-Roza2012,Reilly2013,Reilly2013a}. With the motivation of formulating such a set for HIOSs, we review a set of well-characterized systems of molecules adsorbed on metal surfaces for which detailed information from experiment and calculations exists. We start out with an overview of the major experimental techniques which provide reference data regarding molecular geometry and energetics of adsorbates at surfaces. We then proceed with an overview of the recent advances in DFT-based methodologies with a special focus on density-dependent dispersion-inclusive approaches such as the DFT+vdW$^{\mathrm{surf}}$ method. Following this overview we exemplify interpretation of electronic structure results using PTCDA on Ag(111) data as shown in Fig.~\ref{fig:intro}.  In the following we review and analyze the experimental data for the different systems, which range in complexity from rare-gas adatoms to large conjugated aromatics and carbon nanostructures. Based on this data we attempt a first assessment on the current level of accuracy in first-principles calculations of HIOSs, by using the DFT+vdW$^{\mathrm{surf}}$ method.

\section{Experimental Methods}
\label{chapter-experiment}

Experimental characterization of HIOSs in ultra-high vacuum is performed with a vast set of surface science techniques (see Tab.~\ref{tab-exp}) based on topographic surface imaging, spectroscopy, surface scattering, and thermodynamic measurements, all of which complement each other. 

Imaging techniques such as Transmission Electron Microscopy (TEM)~\cite{Williams1996, Tanishiro1981}, Scanning Tunneling Microscopy (STM)~\cite{Binnig1982, Binnig1983, Tersoff1993} and Atomic Force Microscopy (AFM)~\cite{Binnig1986} play an important role in the initial characterization of adsorbate overlayers at surfaces. In the case of TEM the growth mode, surface reconstruction, the approximate thickness of and phase boundaries between adsorbate overlayers can be measured. Whereas atomic-level resolution in TEM can be achieved by now~\cite{Hansen2001}, structural analysis on the single molecule level is still difficult. Such resolution can, however, be achieved for molecular adsorbates using STM and AFM. This gives the ability to define a model of single molecule adsorbate structure and, in combination with low-energy electron diffraction (LEED)~\cite{Pendry1974, VanHove1979}, also to construct a model of the periodic surface structure and surface unit cell. Recent advances due to tip-functionalization with small molecules~\cite{Schneiderbauer2014} have significantly increased the applicability and lateral resolution of AFM~\cite{Weymouth2014}, which is sometimes even referred to as "sub-atomic resolution AFM"~\cite{Emmrich2015}. Complementary to the information given by these diffraction and imaging techniques, surface spectroscopy methods such as x-ray photoemission (XPS)~\cite{Papp2013}, x-ray absorption spectroscopy (XAS), ultraviolet photoemission (UPS), and Two-Photon Photoemission (2PPE) yield information about the changes in electronic structure along with the nature and extent of the interaction between adsorbate and substrate.

\begin{table}
\caption{\label{tab-exp} List of experimental techniques and their abbreviations (Abbrev.) from which interface structure, adsorbate geometry, and interaction energies can be extracted. }
 \begin{tabular}{ll} \hline \noalign{\vskip 1pt} 
 Overlayer structure, packing, and growth &  Abbrev.  \\ \hline \noalign{\vskip 1pt} 
  Transmission Electron Microscopy & TEM \\
 Scanning Tunneling Microscopy & STM \\
 Atomic Force Microscopy & AFM  \\
 Low-Energy Electron Diffraction & LEED  \\
 Normal Emission Photoelectron Diffraction&  PhD \\
 UV/ X-Ray Photoemission Spectroscopy & UPS/XPS \\ 
 X-Ray Absorption Spectroscopy & XAS \\ 
 High Resolution Electron Energy Loss & HREELS \\  \hline \noalign{\vskip 1pt} 
 Adsorbate Height and Geometry & Abbrev.    \\ \hline \noalign{\vskip 1pt} 
 Near-Incidence X-Ray Standing Waves & NIXSW \\ 
 Near Edge X-Ray Absorption Fine-Structure & NEXAFS \\ 
 Normal Emission Photoelectron Diffraction & PhD \\ \hline \noalign{\vskip 1pt} 
 Adsorption Energy  & Abbrev. \\ \hline \noalign{\vskip 1pt} 
 Temperature Programmed Desorption & TPD  \\
 Single Crystal Adsorption Microcalorimetry & SCAM \\ 
 AFM force pulling experiments &  \\
  \hline 
 \end{tabular} 
\end{table}

The above methods can produce the basic information of overlayer surface unit cell, lateral adsorbate arrangement, and surface chemical shift that often serve as input to construct first-principles models of HIOSs. Using methods based on DFT and many-body perturbation theory (MBPT), stable adsorption geometries can then be calculated, which may or may not support the initial experimental model. Such calculations yield additional  detailed information on the individual atomic positions and the interaction strength between adsorbate and substrate. Whereas the above mentioned experimental techniques do not directly give access to geometry, validation can be achieved indirectly through e.g. the comparison of spectroscopic signatures with chemical core-level shifts and diffraction patterns obtained from an accurate atomistic model of geometry and energetics.

In turn, a relatively novel and powerful tool for validation of atomistic models are experimental techniques that enable an accurate determination of individual atomic adsorption heights and intramolecular structure based on model fitting. The most common such techniques are Near-Incidence X-Ray Standing Wave measurements (NIXSW)~\cite{Zegenhagen1993, Woodruff1994, Woodruff2005}, Normal Emission Photoelectron Diffraction (PhD)~\cite{Woodruff1994a}, and Angular-Resolved Near Edge X-Ray Absorption Fine-Structure (NEXAFS)~\cite{Stohr1992}. In NIXSW an x-ray wave at normal incidence forms a standing wave pattern with its reflection that has the same periodicity as the Bragg planes of the underlying substrate. By tuning the photon energy of incoming x-ray light, the standing wave pattern shifts and the photoelectron spectra can be measured as a function of the x-ray incidence energy. Using Fourier vector analysis, the corresponding atomic positions and the coherence of the signal can be matched with model geometries~\cite{Mercurio2014}. The corresponding structural models are highly accurate with an experimental uncertainty in the range of 0.10~\AA{} in the determination of the vertical heights. The disadvantages of NIXSW, similar to PhD, are the immense complexity of the spectral model fits and the need for near perfect adsorbate overlayer order. While NIXSW and PhD both yield accurate determination of vertical heights, they also provide limited information on internal molecular degrees of freedom, except for inferred information from the model structures used in the spectral fit. In this respect, angular-resolved NEXAFS serves as a powerful complementary technique that enables determination of the relative orientation and angle of chemically-distinct molecular sub-domains with respect to the surface. NEXAFS-based structural characterization has been successfully used for complex systems such as metal-adsorbed porphyrine~\cite{Diller2012,Diller2013} and azobenzene~\cite{Piantek2009,McNellis2010,Gahl2010} derivatives.

%%TPD
The dominant technique employed to determine the interaction energy of adsorbates on surfaces is temperature programmed desorption spectroscopy (TPD)~\cite{King1975}. In a TPD experiment, a sample is slowly heated at a constant rate while monitoring, at the same time, the rate of appearance of gases desorbed from the surface. The measured desorption temperature of a sufficiently diluted adsorbate overlayer ideally reflects the interaction strength of a single molecule on the surface. 
A variety of different techniques exist to analyze the corresponding desorption spectra, all of which are based on the Polanyi-Wigner equation (PWE). Integral methods such as the ones proposed by Redhead~\cite{Redhead1962} or Chan, Aris, and Weinberg~\cite{Chan1978} impose an assumption on the order of the desorption reaction and additionally assume coverage-independence of both the desorption energy and the entropy-related pre-exponential factor. These methods can therefore not be applied to adsorbates exhibiting lateral  interactions~\cite{Jong1990}. Differential techniques to extract interaction energies such as the one proposed by \citet{King1975} or Habenschaden and K\"uppers~\cite{Habenschaden1984} utilize a large set of desorption measurements and make no assumptions on prefactors as function of temperature and coverage. All methods share that they have been devised for the study of small adsorbates of interest at that time, therefore not addressing the characteristics of complex, extended adsorbates exhibiting strong lateral interactions and large configurational freedom~\cite{Fichthorn2002, Fichthorn2007, Campbell2012, Weaver2013, Campbell2013a}. 

Single crystal adsorption microcalorimetry (SCAM)~\cite{Borroni-Bird1991,Borroni-Bird1991a,Stuck1996} is an alternative approach that does not suffer from many of the difficulties that TPD analysis faces. It has the advantage that the observed radiated heat from the surface is directly connected to the adsorption energy by knowledge of the heat capacity. The main disadvantage is the complexity of its experimental setup and calibration, resulting in the operation of only few microcalorimeters at the moment.

One important additional technique to measure the interaction strength between adsorbate and substrate is represented by AFM force pulling experiments of single molecules~\cite{Wagner2012, Wagner2014}. The measured force law of the single molecule pulling event can be related to a model potential with a well-depth that corresponds to a free energy of desorption. This technique also opens the possibility to model different independent interaction contributions that contribute to the force pulling signal. However, there are some unresolved difficulties in relating the integrated binding energy of the force pulling event with interaction energies from first-principles calculations.

A number of spectroscopic techniques could be used as fingerprint methods to the structure and interaction strength of molecules on surfaces. Photoelectron spectroscopy, pump-probe spectroscopy such as 2PPE, and Scanning Tunneling Spectroscopy yield the positions of adsorbate molecular resonances with respect to the Fermi level of the substrate~\cite{Bogner2015}. In combination with MBPT simulations, the image-charge potential-induced state renormalization of molecular level alignment~\cite{Neaton2006} can be used to gauge on the adsorption height of the adsorbate. Surface-enhanced Raman~\cite{Moskovits2013,Gruenke2016} and Sum-Frequency Generation (SFG) spectroscopy are surface science techniques that have recently gained popularity and have been used to determine the layer thickness and structure of molecules on surfaces~\cite{Xu2012,Lee2011}. The corresponding spectral shapes furthermore enable insight into which molecular moieties are strongly or loosely bound to the surface~\cite{Klingsporn2014}. High resolution electron-energy loss spectroscopy (HREELS) provides insight into the vibrational and electronic properties of adsorbed molecules and has been used extensively in combination with simulations to determine adsorbate structure.~\cite{Maass2016,Hahn2015} A significant portion of future method development in surface science will be geared towards combination of existing methods, such as STM tip-enhanced Raman spectroscopy~\cite{Bailo2008}, four-wave mixing spectroscopy~\cite{Renger2010}, or transient spectroscopy during molecular scattering~\cite{Chen2005}.

%AFM
%force pulling stuff, IBM
%TPD and analysis vs. Microcalorimetry
%
%STM, STS
%NIXSW
%NEXAFS, XPS
%Wodtke-Experiments, pump-probe time-resolved, 2PPE in general
%surface-enhanced RAMAN, Sum-Frequency generation
%
%LEED-IV for overlayer structure, Photodiffraction, 

\section{Theoretical Methods}
\label{chapter-theory}

The PTCDA molecule in Fig.~\ref{fig:intro} serves as an ideal benchmark candidate exhibiting all relevant aspects of  adsorbate-substrate interaction. The terminal anhydride-oxygens are chemically bound to the metal surface, whereas the conjugated aromatic core of the molecule induces attractive dispersion interactions between adsorbate and substrate. The relatively small adsorption height also leads to an increased Pauli repulsion of the closed-shell molecule core on the substrate. Finally, the level-alignment of molecular resonances with respect to the Fermi level of the surface determines the amount of charge transfer between adsorbate and substrate. An accurate description of the surface-induced molecular distortion, adsorption height, and interaction strength can only be achieved by accounting for all the effects that we have mentioned above. If we also consider the system size at hand, high computational efficiency becomes an equally important factor. 

DFT, as the electronic-structure method of choice in condensed matter physics, represents a good compromise between accuracy and computational efficiency. However, the above discussed interactions are often not described with sufficient accuracy using current  (semi-)local and hybrid xc approximations to the exact density functional. 
Chemical interactions between adsorbate species and metallic substrates are captured relatively well at the level of the Generalized Gradient Approximation (GGA) of which the functional developed by Perdew, Burke, and Ernzerhof (PBE)~\cite{Perdew1996} is a popular variant, albeit at a tendency to overestimate adsorption energies in strongly interacting systems~\cite{Hammer1999, Feibelman2001}. An accurate description of Pauli repulsion effects and molecular level alignment requires a more sophisticated description of exchange. The simple admixture of exact exchange on the Hartree-Fock level may simply not be sufficient~\cite{Stroppa2007} and in some cases can even lead to an overestimation of the substrate bandwidth and exchange splitting~\cite{Marsman2008}. Several recent works have developed correlation descriptions for solids and surfaces based on the RPA~\cite{Rohlfing2008,Olsen2013} and beyond~\cite{Ren2011, Olsen2012, Olsen2013b} in combination with different variants of (screened) exact exchange~\cite{Ren2011} that promise chemical accuracy for short-range interactions between adsorbates and surfaces~\cite{Schimka2010}. 

Admitting that several challenges remain on the level of short-range correlation and exchange, the biggest challenge in simulating HIOSs is the accurate treatment of non-local correlation effects such as dispersion interactions. Whereas the most straightforward treatment of dispersion interactions follows incorporation of non-local correlation into DFT via the Adiabatic-Connection Fluctuation Dissipation Theorem~\cite{Bohm1953, Gell-Mann1957, DiStasio2014}, it is certainly still limited in terms of efficiency from the computational perspective. A hierarchy of different and more efficient approaches to incorporate long-range dispersion into Density-Functional Approximations (DFAs) exists~\cite{Klimes2012}. These can be grouped into three major categories: (1) empirical \emph{a posteriori} dispersion correction approaches, (2) density-dependent dispersion functionals, (3) and the aforementioned correlation functionals directly based on RPA.

The first category is prominently represented by the series of methods proposed by \citet{Grimme2004, Grimme2006, Grimme2010}. In this case, an existing DFA is complemented by a pairwise-additive correction to the total energy which exhibits the $R^{-6}$ behavior of the leading-order dispersion term based on empirical pretabulated parameters for atomic polarizabilities, dispersion coefficients $C_6$, and van der Waals radii. The connection is achieved by a damping function acting on the vdW contribution at short-range. This pragmatic approach has been applied in the description of intermolecular interactions in gas phase complexes~\cite{Grimme2004} and molecule-surface adsorption~\cite{Tonigold2010}.
Despite its low computational cost and fair accuracy for small molecules, its insufficient response to the local chemical environment and collective response effects along with the absence of higher-order dispersion terms lead to a significant overestimation of interaction energies for molecules at surfaces~\cite{Tonigold2010}.

Van-der-Waals functionals (vdW-DF) are representatives of the second category of density-derived methods~\cite{Dion2004, Klimes2011}. An additional vdW contribution to the Hamiltonian is computed as non-local functional of the electron density by a two-point integral and a given integration kernel. This approach bears considerably more computational demand. However, recent improvements in computational efficiency~\cite{Roman-Perez2009} and performance~\cite{Klimes2010,Klimes2011} lead to a widespread use of approaches such as the vdW-DF-cx~\cite{Berland2014, Berland2014a} and optPBE-vdW~\cite{Klimes2011} functionals. Compared to the first category of methods, recent vdW-DF methods yield a considerably improved description of adsorbate structure and energetics for a number of HIOSs~\cite{Li2012,Carrasco2014}.

\begin{figure}
\includegraphics[width=\columnwidth]{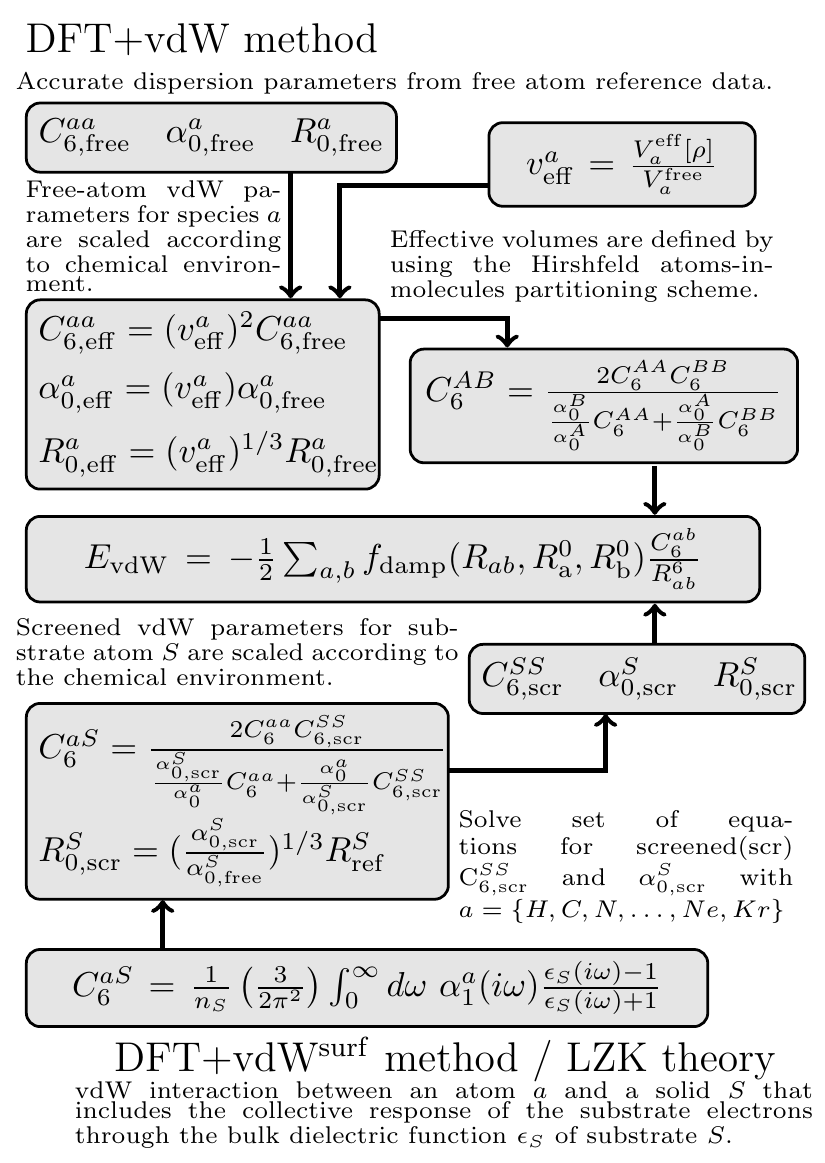}
\caption{\label{fig:vdWsurf} Flowchart explaining the link between Lifshitz-Zaremba-Kohn (LZK) theory and the DFT+vdW method leading to the DFT+vdW$^{\mathrm{surf}}$ method.}
\end{figure}

%vdW and vdW-surf
Several efficient approaches are based on a connection between a pairwise dispersion model and the electron density, namely by constructing vdW parameters such as atomic C$_6$ coefficients, vdW radii $R$, and static atomic polarizabilities $\alpha_0$ as functionals of the chemical environment and the electron density. Such methods include the DFT+XDM approach originally developed by Becke and Johnson~\cite{Becke2005, Kannemann2010, Steinmann2011} and the DFT+vdW approach of Tkatchenko and Scheffler~\cite{Tkatchenko2009}. In the latter approach the vdW parameters  C$_6^a$, $\alpha_0^a$, and $R_{\mathrm{0}}^a$ for an atom $a$ are constructed from free-atom reference data and renormalized by the change in effective volume of the atom in the molecule. The latter effect accounts for the changes in local polarizability and chemical environment of the atomic species (see Fig.~\ref{fig:vdWsurf}, top). The resulting C$_6$ coefficients show a mean absolute relative error (MARE) of 5.5\% for intermolecular C$_6$ coefficients between a variety of atoms and molecules in gas phase.~\cite{Tkatchenko2009} The effective atomic volumes are directly derived from the density using the atoms-in-molecules scheme proposed by Hirshfeld~\cite{Hirshfeld1977}. This approach has been recently modified to extract dispersion parameters directly from charge analysis enabling its use for semi-empirical and tight-binding approximations to DFT~\cite{Stoehr2016}. Furthermore, it should be emphasized that the methods based on the Tkatchenko-Scheffler approach (DFT+vdW, DFT+vdW$^{\mathrm{surf}}$, and DFT+MBD) are proper functionals of the electron density and, hence, should and have been implemented self-consistently in the context of DFT~\cite{FerriPRL2015}. While the self-consistency of the vdW energy can modify electronic properties of solids and surfaces~\cite{FerriPRL2015}, its impact on structures and stabilities is typically minimal (on the order of 0.001 {\AA} and few meV, respectively). For this reason, the calculations in this manuscript were carried out without accounting for self-consistency in the vdW energy.

However, direct application of the above mentioned approaches to molecules at metal surfaces leads to a significant overestimation of the adsorbate-substrate interaction as has been observed for example for azobenzene and its derivatives adsorbed at coinage metal surfaces~\cite{McNellis2009,Mercurio2010, Schulze2014}. The recipe behind the DFT+vdW method of rescaling accurate free-atom reference vdW parameters according to the chemical environment of each atomic component often yields a highly accurate description of dispersion interactions between atoms and molecules in gas phase. However, the non-local correlation interaction between adsorbate atoms and an extended metal surface requires account of the collective many-body substrate response rather than only the local atom-atom response of individual metal atoms with the atoms of the adsorbate.~\cite{Zaremba1976} The DFT+vdW$^{\mathrm{surf}}$ method~\cite{Ruiz2012} accounts for this by modelling screened vdW interactions in the adsorption of atoms and molecules on metal surfaces. On the basis of the Lifshitz-Zaremba-Kohn (LZK) theory~\cite{Lifshitz1956,Zaremba1976} and its equivalent formulation in terms of interatomic pairwise potentials,~\cite{Bruch1997,Patil:Tang:ToenniesJChemPhys2002} the collective effects of the atom-substrate interaction are projected onto renormalized C$_6^{aS}$ coefficients that describe the dispersion interaction between adsorbate atoms $a$ and substrate atoms $S$. These coefficients are expressed in terms of an integral over the frequency-dependent polarizability $\alpha_1^a(i\omega)$ of the adsorbate atom and the dielectric function $\epsilon_S$ of the substrate (see Fig.~\ref{fig:vdWsurf}, bottom). With this formulation, the C$_6^{aS}$ coefficients effectively ``inherit'' the collective effects contained in the many-body response of the solid. Obviously, not all many-body effects of the extended surface can be treated in this effective way. While the vdW$^{\mathrm{surf}}$ method exactly reproduces the long-range vdW energy limit by construction, many-body effects closer to the surface are included approximately utilizing the electron density. Accurately treating \textit{all} many-body effects in the vdW energy would require fully non-local microscopic approaches to the correlation energy, such as those based on the adiabatic connection formalism~\cite{Bohm1953, Gell-Mann1957, DiStasio2014}.
Finally, the vdW$^{\mathrm{surf}}$ parameters for a given substrate species are calculated using the combination rule of the vdW method and solving C$_6^{SS}$ and $\alpha_0^S$ with a linear set of equations for a number of different adsorbate species (see Fig.~\ref{fig:vdWsurf}, bottom).~\cite{Tkatchenko2009} The resulting effective DFT+vdW$^{\mathrm{surf}}$ scheme has been applied to numerous HIOSs~\cite{Liu2012, Mercurio2013, Ruiz2012, Schuler2013, Burker2013,Liu2013,Liu2014} yielding adsorption geometries that are in good agreement with experiment as we will detail in the following chapters.

The DFT+vdW$^{\mathrm{surf}}$ method introduced above is generally applicable to model adsorption on solids, independent of whether they are insulators, semiconductors, or metals. The LZK theory is an exact asymptotic theory for any polarizable material and the firm foundation of vdW$^{\mathrm{surf}}$ on the LZK theory ensures its transferability. The vdW$^{\mathrm{surf}}$ approach necessitates the dielectric function of the bulk solid as an input, which can be calculated from time-dependent DFT, RPA, or taken from experimental measurements. However, the method is applicable to surfaces with any termination, defects, and other imperfections, because the vdW parameters depend on the electron density at the interface. The transferability of the vdW$^{\mathrm{surf}}$ method to different surface terminations has been recently demonstrated for adsorption on (111), (110), and (100) metallic surfaces~\cite{Ruiz:Liu:Tkatchenko2016}.

Despite this success, several challenges remain to model dispersion interactions in HIOSs accurately. While the description of adsorption geometries seems adequate at the level of effective pairwise interactions, adsorption energies still appear systematically overestimated. This is due to the missing beyond-pairwise interactions and the neglect of the full many-body response of the combined adsorbate-substrate system~\cite{Tkatchenko2012, Maurer2015}. The recently developed DFT+MBD method tackles this problem by calculating the full long-range many-body response in the dipole limit~\cite{Tkatchenko2012,Tkatchenko2013,DiStasio2014}. In short, the MBD method makes an approximation to the density-density response function, consisting of a set of atom-centered interacting quantum harmonic oscillators. Under this employed assumption, the MBD method is equivalent to RPA. Initial results for molecules on metal surfaces, which include the Xe atom, benzene~\cite{Liu2015}, PTCDA, and graphene on metal surfaces~\cite{Maurer2015}, are promising. However, the current MBD approach has been developed to describe the correlation problem for molecules and finite-band gap materials with atom-centered quantum harmonic oscillators, which would not fully account for the delocalized plasmonic response of free electrons in the metal substrate. Despite several remaining questions, a systematic improvement of the current MBD scheme is possible, potentially opening a path towards the exact treatment of dispersion energy at drastically reduced computational cost.

All the DFT calculations presented herein are performed employing the PBE+vdW$^{\mathrm{surf}}$ functional, by means of the full-potential all-electron code \textsc{fhi-aims}~\cite{blum2009ab, havu2009efficient} and the periodic plane wave code \textsc{CASTEP}~\cite{Clark2005}.

\section{Interpretation of Electronic Structure Calculation Results}
\label{chapter-interpretation}

\begin{table}
\caption{\label{tab-theo} List of electronic structure methods and their abbreviations (Abbrev.). }
\begin{tabular}{ll} \hline \noalign{\vskip 1pt} 
Density-Functional Approximations & Abbrev. \\ \hline \noalign{\vskip 1pt}
Local Density Approximation  & LDA \\
Generalized Gradient Approximation & GGA \\
GGA by \citet{Perdew1996} & PBE  \\ \hline
Semi-empirical dispersion methods & Abbrev. \\ \hline \noalign{\vskip 1pt} 
Grimme dispersion correction~\cite{Grimme2004} & DFT-D \\
Grimme dispersion correction 2nd gen.~\cite{Grimme2006} & DFT-D2 \\
Grimme dispersion correction 3rd gen.~\cite{Grimme2010} & DFT-D3 \\ \hline \noalign{\vskip 1pt} 
Density-derived dispersion methods & Abbrev. \\ \hline \noalign{\vskip 1pt} 
Becke-Johnson method~\cite{Becke2005, Kannemann2010, Steinmann2011} & DFT+XDM \\ 
Tkatchenko-Scheffler (TS) method~\cite{Tkatchenko2009} & DFT+vdW \\ 
TS incl. collective metal response~\cite{Ruiz2012} & DFT+vdW$^{\mathrm{surf}}$ \\ 
many-body dispersion method~\cite{Tkatchenko2012} & DFT+MBD \\ 
Non-local functionals~\cite{Klimes2010, Klimes2012} & vdW-DF \\
optimized exchange vdW-DF~\cite{Klimes2011} & optPBE,optB86b \\ 
consistent exchange vdW-DF~\cite{Berland2014,Berland2014a} & vdW-DF-cx  \\ \hline \noalign{\vskip 1pt} 
many-body correlation methods & Abbrev. \\ \hline \noalign{\vskip 1pt} 
Exact Exchange~\cite{Hesselmann2007} & EX \\
Random Phase Approximation~\cite{Rohlfing2008,Olsen2013} & RPA or cRPA \\ \hline
 \end{tabular} 
\end{table}

Having summarized the ingredients of many existing dispersion-inclusive electronic structure methods (see Table~\ref{tab-theo}), we revisit the adsorption of PTCDA on Ag(111), as depicted in Fig.~\ref{fig:intro}. We do this to illustrate the wide range of interactions that need to be accounted for by a first-principles method to accurately describe HIOSs and to establish the merits of different types of methods in direct comparison to experiment.

In the case of PTCDA on Ag(111) an accurate measurement of the adsorption height from NIXSW exists, with 2.86~\AA{} for the average height of the carbon backbone~\cite{Hauschild2010}. The value of adsorption energy remains to be directly measured. Several disputed estimates for the single-molecule and monolayer adsorption energy are given in literature, ranging from 1.40 to 3.46~eV~\cite{Ruiz2012, Maurer2015,Berland2014a}. 
The value given in Fig.~\ref{fig:intro}, extrapolated from desorption measurements of the smaller homologous NTCDA molecule~\cite{Ruiz2012}, coincidentally corresponds to the median of these estimates. 

A DFT calculation based on a local description of exchange and correlation effects (LDA) underestimates the binding distance, but leads to a seemingly good description of adsorption energy. It is important to note that this apparent agreement stems from an incorrect balance between short-range kinetic, electrostatic, and xc contributions~\cite{TkatchenkoPRB2006}, and that LDA does not include any long-range dispersive interactions. DFAs based on a semi-local xc-description, namely GGAs, such as PBE~\cite{Perdew1992}, result in the opposite extreme case: The functional also lacks any description of long-range correlation, but offers a better description of covalent bonding contributions. The result is negligible binding to the surface as indicated by a large overestimation of adsorption height and underestimation of adsorption energy.

%disp corr.
This insufficient description of long-range correlation makes GGAs an ideal starting point to incorporate dispersion interactions. Pairwise dispersion interaction methods built on-top of GGAs (PBE-D and PBE+vdW) yield bound structures in the range of 2.9 to 3.2~\AA{}, but systematically overestimate the binding energy. Long-range correlation functionals of the vdW-DF family (Fig.~\ref{fig:intro} shows a so-called vdW-DF1~\cite{Dion2004,Berland2014a}) yield a wide range of results depending on the treatment of long-range correlation, long-range exchange and short-range exchange. Depending on the construction, in many cases vdW-DF can yield a good description of either adsorption geometry or adsorption energy.

%vdWsurf
PBE+vdW$^{\mathrm{surf}}$ introduces substrate-screening effects into the PBE+vdW functional and thereby improves the adsorption properties considerably. However, the screened interactions also reduce the effective vdW radii and result in reduced adsorption heights that, as will be shown in the remainder of this work, are in excellent agreement with experiment for a variety of systems. The PBE+vdW$^{\mathrm{surf}}$ enables this at negligible computational overhead compared to PBE.

%MBD and correlation
For some of the larger molecules we discuss in this work, PBE+vdW$^{\mathrm{surf}}$ seems to overestimate the adsorption energy with respect to the limited available experimental reference data. As discussed above, this can be remedied with methods that explicitly account for the many-body nature of dispersion interactions, such as the PBE+MBD method~\cite{Tkatchenko2012,DiStasio2014, Maurer2015} or exact correlation treatment in the Random Phase Approximation (EX+cRPA)~\cite{Rohlfing2008}. Both of these methods yield further reduced adsorption energies and in the case of PBE+MBD excellent agreement with experiment was recently established for benzene~\cite{Liu2015} and azobenzene~\cite{Maurer2016} on Ag(111). Published EX+cRPA results for PTCDA on Ag(111) yield a larger deviation from the experimental reference, however, they did not account for geometrical relaxation and a full numerical convergence of RPA remains a challenging issue~\cite{Rohlfing2008}. 

The presented case of PTCDA on Ag(111) remains somewhat disputed due to the lack of a directly measured experimental adsorption energy. However, this example nicely illustrates the importance of unambiguous experimental reference values in the development of improved electronic structure methodologies and motivates the development of the here presented benchmark dataset.

\section{Overview of Benchmark Systems}
\label{chapter-overview}

\begin{figure}[htp]
\centering
\includegraphics{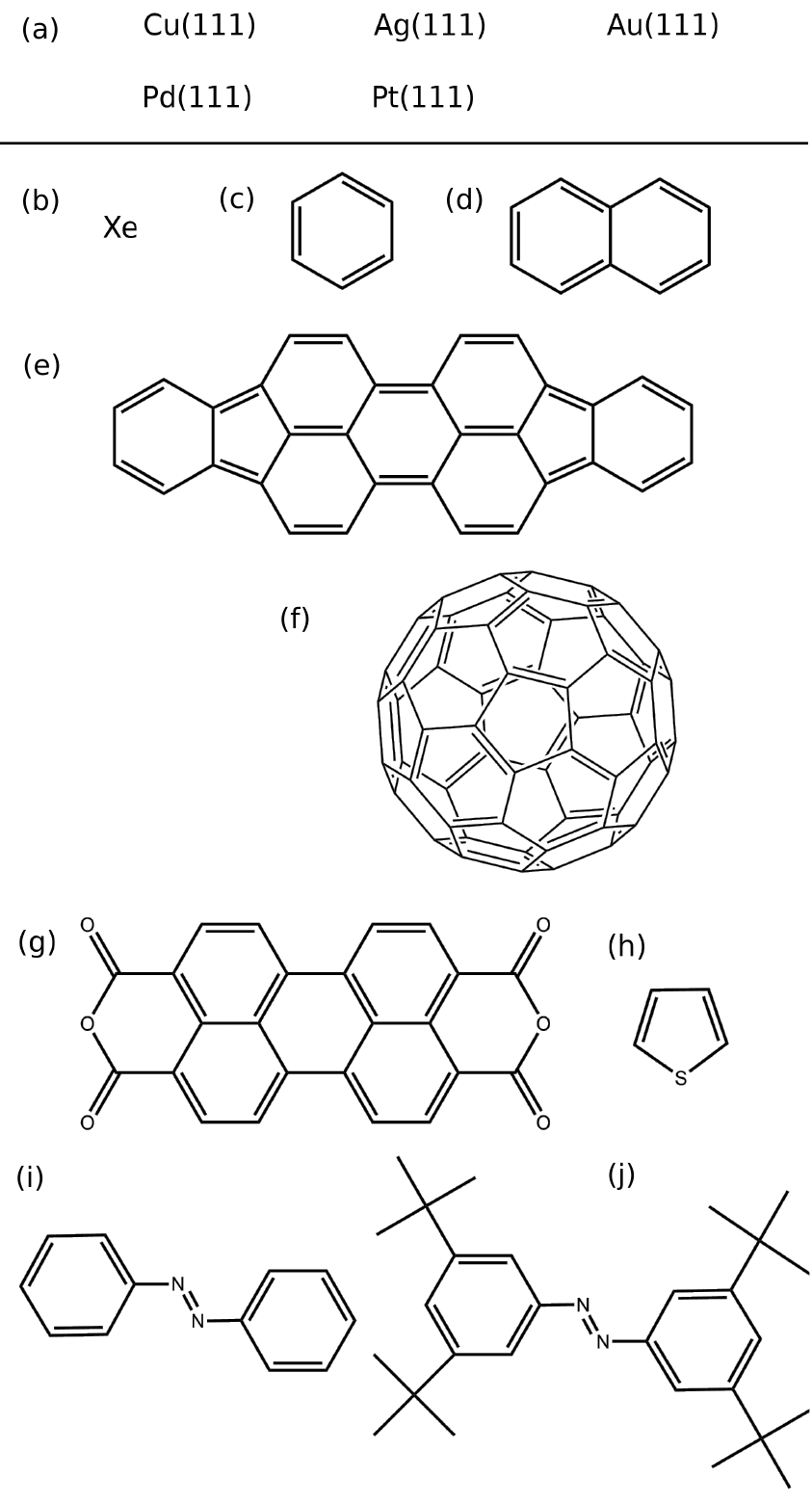}
\caption{\label{fig-summary} Summary of surfaces and molecules incuded in the benchmark set: (a) 7 close-packed transition metal surfaces, (b) Xenon, (c) benzene (Bz), (d) naphthalene (Np), (e) Diindenoperylene (DIP), (f) C$_{60}$ Buckminster-Fullerene, (g) 3,4,9,10-perylene-tetracarboxylic acid (PTCDA), (h) Thiophene (Thp), (i) E-Azobenzene (AB), (j) E-3,3',5,5'-tetra-\emph{tert}-butyl-Azobenzene (TBA) }
\end{figure}

In the following we will review the adsorption properties of different classes of organic compounds (see Table~\ref{tab-systems}) on metal surfaces. From each of these classes we will select a number of test cases representing different limiting cases of adsorbate-substrate interaction. Fig. \ref{fig-summary}(a) depicts different metal substrates for which the interaction with exemplary adsorbates (Fig. \ref{fig-summary}(b) to (j)) is considered. In each case we will review the experimental reference data and the calculated geometry and energetics as predicted by DFT+vdW$^{\mathrm{surf}}$.

\begin{table}
\caption{\label{tab-systems} List of molecules and their abbreviations (Abbrev.) included in the benchmark set. }
 \begin{tabular}{ll} \hline \noalign{\vskip 1pt} 
Atom/Molecule & Abbrev. \\ \hline \noalign{\vskip 1pt} 
Xenon & Xe \\
Benzene & Bz \\
Napthalene & Np \\ 
Diindenoperylene & DIP \\ 
(C$_{60}$-I$_h$)[5,6]fullerene  & C$_{60}$ \\ 
3,4,9,10-perylene-tetracarboxylic acid & PTCDA \\ 
Thiophene & Thp \\ 
E-Azobenzene, (E)-Di(phenyl)diazene & AB \\
E-3,3',5,5'-tetra-\emph{tert}-butyl-Azobenzene & TBA \\ \hline
 \end{tabular} 
\end{table}

We group the set of hybrid organic-metal benchmark systems into rare-gas adsorption at coinage metals on the example of Xe atom, aromatic compounds adsorbed at metal systems, extended and compacted carbon nanostructures, Sulfur-containing compounds represented by thiophene on Cu(111), Ag(111), and Au(111), Oxygen-containing compounds represented by 3,4,9,10-perylene-tetracarboxylic acid (PTCDA) adsorbed at coinage metal surfaces, and Nitrogen-containing compounds represented by E-azobenzene and E-3,3',5,5'-tetra-\emph{tert}-butyl-azobenzene (E-TBA) adsorbed at Ag(111) and Au(111).

%%%%%%%%%%%%%%%%%%%%%%%%%%%%%%%%%%%%%%%%%%%%%%%%%%%
\section{Rare-Gas Adsorption on Metal Surfaces}
\label{chapter-raregas}
\paragraph{Experimental Data}

The adsorption of noble gases on metal surfaces has been extensively studied as prototypical example of physisorption. A historical perspective of these studies can be found in the works, for example, by \citet{Diehl:Seyller:Caragiu:etalJPCondensMatter2004} and \citet{DaSilvaPhDthesis}. An exhaustive historical survey is out of the scope of this work. We will restrict ourselves to experimental data on structures and adsorption energetics. Moreover, we focus here exclusively on the adsorption of Xe on transition-metal surfaces. From this perspective, the most important fact is the paradigm shift that occurred 25 years ago with respect to the preferred adsorption site of Xe. The general assumption prior to 1990 was that the adsorption potential of noble gases on surfaces would be more attractive in high-coordination sites than in those with lower coordination. In the case of Xe, for instance, experimental studies using spin-polarized LEED suggested the hollow site as the preferred adsorption site on close-packed metal surfaces \cite{Potthoff:etal:SS1995,Hilgers:etal:SS1995}. This changed with the dynamical LEED studies of adsorbed Xe on Ru(0001)~\cite{Narloch:Menzel:CPL1997}, Cu(111)~\cite{Seyller:Caragiu:Diehl:etalCPL1998}, Pt(111) \cite{Seyller:Caragiu:etalPRB1999}, and Pd(111)~\cite{Caragiu:Seyller:DiehlPRB2002}; which showed that Xe atoms reside on top of the substrate atoms instead of in higher-coordination sites~\cite{Diehl:Seyller:Caragiu:etalJPCondensMatter2004}. The other important experimental finding is that Xe adopts a $ (\sqrt{3} \times \sqrt{3})\mathrm{R30}^{\circ}$ structure on Cu(111), Pt(111), and Pd(111). In the case of Xe on Cu(110), a $(12 \times 2)$ structure is formed at low temperature, which consists of rows of adatoms that are commensurate with the substrate, having higher-order commensurate periodicity along the substrate rows of the surface and a spacing between the rows that is equal to the Cu row spacing~\cite{Caragiu:Seyller:DiehlSurfSci2003,Diehl:Seyller:Caragiu:etalJPCondensMatter2004}. Most importantly, the LEED studies by \citet{Caragiu:Seyller:DiehlSurfSci2003} indicate that the Xe rows are located on top of the Cu substrate rows. 

For this work, we take the review papers by \citet{Diehl:Seyller:Caragiu:etalJPCondensMatter2004} and \citet{Vidali:Ihm:Kim:etalSSRep.1991} as our guidelines for the experimental data on the adsorption of Xe on transition-metal surfaces. The overview of these data is shown in Tab.~\ref{Table_XeMe111_Results} while the details of each experiment can be found in the original references. The experimental adsorption distances in these systems were mainly obtained using the LEED technique. The experimental adsorption energies are mostly a result of TPD. These experiments report exponential prefactors of desorption of the order of $10^{12}-10^{13}$ s$^{-1}$, which are in the expected range for simple adsorbates and small molecules~\cite{Fichthorn2002,Tait:etalJChemPhys2005}.
%%%%%%%%%%
\begin{table*}[ht]
\begin{center}
\caption{Adsorption distances and energies for Xe on transition-metal surfaces in \AA{} and eV respectively.
Both adsorption distances and energies correspond to the system after relaxation. The values of $d$ and $E_{\mathrm{ad}}$ for Ag(111) correspond to the best estimates in Ref.~\onlinecite{Vidali:Ihm:Kim:etalSSRep.1991}. The experimental data is taken from Refs.~\onlinecite{Diehl:Seyller:Caragiu:etalJPCondensMatter2004,Vidali:Ihm:Kim:etalSSRep.1991,Seyller:Caragiu:Diehl:etalCPL1998,Caragiu:Seyller:DiehlSurfSci2003,Pouthier:Ramseyer:Girardet:etalPRB1998,Zhu:Ellmer:Malissa:etalPRB2003,Caragiu:Seyller:DiehlPRB2002,Hilgers:etal:SS1995,Widdra:Trischberger:etalPRB1997,Seyller:Caragiu:etalPRB1999,Bruch:Graham:Toennies:MolPhys1998,Hall:Mills:etalPRB1989,Braun:Fuhrmann:etalPRL1998,Zeppenfeld:Buechel:etalPRB1994,Ramseyer:Pouthier:Girardet:etalPRB1997}, and~\onlinecite{Gibson:SibenerJChemPhys1988}.}
\label{Table_XeMe111_Results}
\begin{tabular}{ccccccccccc}
& \multicolumn{2}{c}{$ d $ [\AA]} & \multicolumn{2}{c}{$ E_{\rm{ad}} $ [meV]}  \\
& PBE+vdW$^{\mathrm{surf}}$ & Exp. & PBE+vdW$^{\mathrm{surf}}$ & Exp. \\
 \hline \noalign{\vskip 2pt} 
Xe/Pt(111) & 3.46 & $3.4 \pm 0.1$  &  $254$  & $260$ -- $280$ \\
Xe/Pd(111) & 3.12 & $3.07 \pm 0.06$ &  $276$  & $310$ -- $330$ \\
Xe/Cu(111) & 3.46 & $3.60 \pm 0.08$ &  $248$  & $173$ -- $200$ \\
Xe/Cu(110) & 3.29 & $3.3 \pm 0.1$  &  $249$  & $212$ -- $224$ \\
Xe/Ag(111) & 3.57 & $3.6 \pm 0.05$ &  $237$  & $196$ -- $226$ \\
\hline
\end{tabular}
\end{center}
\end{table*}

\paragraph{Theoretical Data}
Even if the experiments have identified the low-coordination top site as the preferred adsorption site for Xe on transition-metal surfaces, they have not been able to identify the origin of this preference. It is in this regard that first-principles calculations have the potential to contribute to the atomistic understanding of the origin of this preference. For an extensive review of first-principles simulation of rare gas adsorption, we point the interested reader to the works of \citet{Diehl:Seyller:Caragiu:etalJPCondensMatter2004}, \citet{DaSilvaPhDthesis}, and \citet{Chen:Al-Saidi:JohnsonJPCM2012}. The predominant physisorptive character of the binding makes rare-gases on metal surfaces ideal test systems for dispersion-inclusive DFT. In general, before the advent of several vdW-inclusive DFT based methods in the last years, the LDA has been used extensively to study the adsorption of Xe on metals~\cite{MuellerPRL1990,DaSilva:Stampfl:SchefflerPRL2003,DaSilva:Stampfl:SchefflerPRB2005}, where it has been found that the top site is energetically more stable by, at most, 50 meV with respect to the hollow adsorption site. In addition, \citet{DaSilva:StampflPRB2008} also studied the adsorption of additional rare gases on metal surfaces using GGAs as xc functional, where they found that these other rare gases also prefer the top adsorption site with the exception of Ar and Ne on Pd(111)~\cite{DaSilva:Stampfl:SchefflerPRB2005,Chen:Al-Saidi:JohnsonJPCM2012}. In general, \citet{Chen:Al-Saidi:JohnsonJPCM2012} mention that GGA xc functionals tend to underestimate the adsortion energy of these systems by a great margin whereas LDA yields equilibrium adsorption distances that are too short in comparison to experiments. 

In this respect, it is currently well established that GGA functionals such as PBE cannot describe systems that are dominated by vdW interactions in an accurate manner. The studies performed by \citet{Chen:Al-Saidi:JohnsonPRB2011,Chen:Al-Saidi:JohnsonJPCM2012} report the performance of several vdW-inclusive DFT methods, such as vdW-DF, vdW-DF2, and DFT-D2, on the adsorption of noble gases on metal surfaces. We have also analyzed the structure and stability of the adsorption of Xe on selected transition metal surfaces with the PBE+vdW and PBE+vdW$^{\mathrm{surf}}$ methods taking into consideration that the latter includes the collective response of the substrate electrons in the determination of the vdW contribution \cite{Ruiz:Liu:Tkatchenko2016}. We reproduce the PBE+vdW$^{\mathrm{surf}}$ results for the top adsorption site of Xe on transition metal surfaces in Tab.~\ref{Table_XeMe111_Results}. 
Before proceeding to discuss these results, we define here the adsorption distance, $d$, as the distance between the vertical coordinate of the Xe atom with respect to the position of the unrelaxed topmost metal layer.
This definition allows us to compare our data with results from NIXSW experiments.
In case of other experimental sources, \textit{e.g.} LEED, $d$ is computed considering the relaxed metal surface.
The adsorption energy, $E_{\rm ad}$, of a system is computed \textit{via} the general definition:
\begin{equation}\label{ads_energy}
E_{\rm ad}=\frac{1}{N} \left[ E_{\rm AdSys}-(E_{\rm Me}+E_{\rm Mol}) \right]\,,
\end{equation}
where $E_{\rm AdSys}$ denotes the energy of the system, while $E_{\rm Me}$ and $E_{\rm Mol}$ refer respectively to the energies of the clean substrate and the molecule in gas phase.
$N$ corresponds to the total number of molecules in the unit cell.

With respect to adsorption site preference, we have found that both adsorption sites, top and fcc-hollow, are almost energetically equivalent using the PBE+vdW$^{\mathrm{surf}}$ method. The top adsorption site is energetically favored by approximately 5 meV for Pd(111) and Ag(111), and 10 meV for Cu(110). Both adsorption sites are virtually degenerate within our calculation settings in the cases of Pt(111) and Cu(111). Nevertheless, the differences in energy between adsorption sites are too small, just a few meV, to regard them as definitive. This same fact has also been found recently by \citet{Chen:Al-Saidi:JohnsonPRB2011}, who reported a few meV difference in their vdW-DF2 calculations between top and fcc-hollow adsorption sites. They have suggested that experimental results cannot be explained by energy differences between top and fcc-hollow adsorption sites. Instead, by examining the two-dimensional potential energy surface (PES) of Xe on Pt(111), they found that the fcc-hollow adsorption sites correspond to local saddle points in the PES, while top sites correspond to a true minimum. Hence, fcc-hollow sites are transient states and thus not easily observed in  experiments.~\cite{Chen:Al-Saidi:JohnsonPRB2011,Chen:Al-Saidi:JohnsonJPCM2012,Bruch1997} This result is general, according to their calculations, for the adsorption of noble gases on transition metal surfaces. This result is further reported to be independent of the underlying xc functional.

In Tab.~\ref{Table_XeMe111_Results} we show the adsorption distances and energies with the PBE+vdW$^{\mathrm{surf}}$ method for the top adsorption site along with available experimental results. In general, the calculated adsorption distances are within 0.10~\AA{} of the experimental results except for Xe on Cu(111), in which the agreement is within 0.14~\AA{} of the experimental value. We have not found significant differences in the adsorption distance of these systems between PBE+vdW and PBE+vdW$^{\mathrm{surf}}$ calculations with the exception of Xe on Cu(110), in which the distance predicted by the PBE+vdW method is 0.12~\AA{} shorter than the PBE+vdW$^{\mathrm{surf}}$ result (see Ref.~\onlinecite{Ruiz:Liu:Tkatchenko2016} for more details).  On the other hand, we have also found that the PBE+vdW$^{\mathrm{surf}}$ results are in closer agreement (within 0.10~\AA) to experimental results than those calculated with other vdW-inclusive DFT methods~\cite{Chen:Al-Saidi:JohnsonPRB2011,Chen:Al-Saidi:JohnsonJPCM2012}. Tab.~\ref{Table_XeMe111_Results} also shows that the PBE+vdW$^{\mathrm{surf}}$ adsorption energies are in good agreement with experimental results. These calculations slightly underestimate the adsorption energy in the cases of Pt(111) and Pd(111), while slightly overestimating it in the case of both Cu surfaces and Ag(111). Nevertheless, these discrepancies amount to about 50~meV out of the range of experimental results in the worst case. The influence of screening is more noticeable in the computation of the adsorption energy. Neglecting the effects of the collective response of the solid leads to an overestimation of the adsorption energy as a result of the inexact magnitude of the energetic contribution originated in vdW interactions. We have exemplified this effect in the adsorption of Xe on metal surfaces with PBE+vdW calculations in Ref~\onlinecite{Ruiz:Liu:Tkatchenko2016}.  

\section{Aromatic Molecules Adsorbed on Metal Surfaces}
\label{chapter-aromatics}

\subsection{Benzene on Cu(111), Ag(111), Au(111)}
\begin{figure}
\includegraphics[width=\columnwidth]{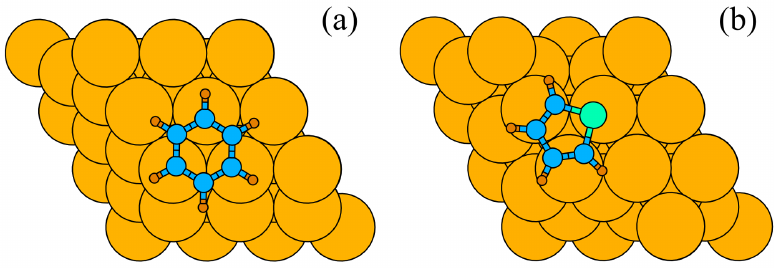}
\caption{\label{fig-bz-thioph} (a) Bz adsorbed on hcp hollow site of Au(111). (b) Thiophene adsorbed on hcp hollow site of Au(111).}
\end{figure}

\paragraph{Experimental Data}

From HREELS and NEXAFS studies it was concluded that Bz binds parallel to the Cu(111) surface~\cite{xi1994}. A flat-lying geometry was also reported for Bz on Ag(111) using NEXAFS~\cite{yannoulis1987}. By means of angle-resolved photoemission spectroscopy (ARPES) in combination with LEED, \citet{dudde1990} concluded that Bz molecules are centered around the three-fold hollow sites of the Ag(111) surface, however, they could not clearly identify whether these are fcc or hcp hollow sites.

The interactions of Bz with coinage metal surfaces are significantly weaker than with other transition metals, such as Pt, Pd, Ir, and Rh, since the \textit{d}-band centers of Cu(111), Ag(111), and Au(111) are well below the Fermi level~\cite{liu2013structure}. STM experiments observed that Bz molecules can easily diffuse over the Cu(111) and Au(111) surface at low temperatures, suggesting a flat PES in both cases~\cite{abad2009,mantooth2007,stranick1995}. TPD experiments revealed that at a heating rate of 4~K/s, Bz desorbs from Cu(111) at a low temperature of 225~K~\cite{xi1994}, from Ag(111) at 220~K~\cite{zhou1990}, and from Au(111) at low coverage of 0.1~ML at 239~K~\cite{syomin2001}. In the case of Bz/Ag(111), a combined NIXSW and TPD study recently reported a vertical adsorption height of benzene of $3.04 \pm 0.02$~\AA{} and an adsorption energy of $0.68 \pm 0.05$~eV~\cite{Liu2015}. TPD-basd experimental adsorption energies for Cu(111) and Au(111) are 0.71~\cite{xi1994} and 0.76~eV~\cite{ihm2004}, respectively. More recent results based on the complete analysis method are $0.69 \pm 0.04$~eV and $0.65 \pm 0.03$~eV for Bz/Cu(111) and Bz/Au(111), respectively.~\cite{Liu2015}

\paragraph{Theoretical Data}
Using a (3 $\times$ 3) supercell, we have explored the PES of a single Bz molecule on close-packed coinage metal surfaces~\cite{liu2013structure}.
For Bz on Cu(111), PBE+vdW$\rm^{surf}$ predicts the most stable adsorption site to be the hcp site where the molecule is rotated by $30^{\circ}$ with respect to the high symmetry directions (hcp30$^{\circ}$ site). However, we also only find a small energy corrugation between all adsorption sites.
The average C--Cu adsorption height with PBE+vdW$\rm^{surf}$ at this site is 2.79~{\AA}. The calculated adsorption energy is 0.86~eV, which deviates by 0.15~eV from experiment~\cite{xi1994,2001Lukas}.

The relaxed Bz molecule adsorbs in a flat-lying geometry on the Ag(111) surface, which is consistent with the observations and conclusions from NEXAFS, EELS, and Raman spectroscopy~\cite{dudde1990,yannoulis1987}.
The carbon--metal distance is larger for Ag than for Cu, which suggests a weaker interaction for the former.
This is in agreement with TPD experiments which showed that the Bz molecule desorbs at a lower temperature from Ag(111) than from Cu(111)~\cite{zhou1990,xi1994,2001Lukas}.
PBE+vdW$\rm^{surf}$ predicts a Bz/Ag(111) adsorption energy of 0.75~eV, which is in good agreement with TPD experiments (0.69~eV)~\cite{zhou1990}.
The flatness of the PES for Bz on Au(111) (see the structure in Fig.~\ref{fig-bz-thioph}(a)) which results from our calculations with the PBE+vdW$\rm^{surf}$ method confirms the STM observations that Bz molecules are mobile over the Au(111) terraces even at 4~K~\cite{mantooth2007}.
An almost identical adsorption energy is found for all sites,
which indicates a small barrier for surface diffusion of Bz on the Au(111) surface.
The PBE+vdW$\rm^{surf}$ adsorption energy for Bz/Au(111) (0.74~eV) is in excellent agreement with the TPD experiments (0.76~eV)~\cite{syomin2001}. The reduced agreement in the case of Bz/Cu(111) compared to the fair agreement for the other substrates might point to remaining discrepancies in the structural model of Bz surface adsorption on Cu(111).

\subsection{Benzene on Pt(111)}

\paragraph{Experimental Data}
Bz adsorbed at the Pt(111) surface, including its adsorption and dehydrogenation reactions, is the best studied system among the Bz adsorption systems.
Nevertheless, even the preferred adsorption site remains controversial in experiments. The diffuse LEED intensity analysis suggested that the bridge site with the molecule $30^{\circ}$ tilted with respect to the high symmetry sites (bri30$^{\circ}$) is the most stable site for Bz chemisorbed on the Pt(111) surface~\cite{wander1991}; whilst nuclear magnetic resonance (NMR) results revealed that Bz molecules are located at the atop site~\cite{tirendi1992}.
Inferred from the orientations of the STM images, the coexistence of Bz molecules at both the hcp and fcc sites was concluded~\cite{weiss1993}.
Despite the ambiguous adsorption site, all experiments clearly concluded that the adsorbate lies flat on the Pt(111) surface, binding with the Bz $\pi$ orbitals to the Pt \textit{d} bands.
STM topographs suggest immobile Bz molecules adsorbed on Pt(111), which points to  strong binding at this surface~\cite{weiss1993,stranick1994}.
The Bz molecules are found to adsorb as intact molecule on the Pt(111) surface at 300~K~\cite{ihm2004}. However, for coverages below 0.6~ML, Bz dissociates completely into hydrogen gas and adsorbed graphitic carbon upon heating and fragment desorption is observed~\cite{ihm2004}.
Therefore, microcalorimetric measurements, rather than desorption-based methods (such as TPD, molecular beam relaxation spectroscopy (MBRS), and equilibrium adsorption isotherms), are required to determine the heat of adsorption for Bz on the Pt(111) surface. The corresponding single crystal adsorption calorimetry results by \citet{ihm2004} report a zero-coverage extrapolated adsorption energy of 2.04~eV.

\paragraph{Theoretical Data}
For Bz on Pt(111), the PBE+vdW$\rm^{surf}$ method predicts the bri30$^{\circ}$ site as the most favorable site with an adsorption energy of 1.96~eV. The second and third preferred sites are the hcp0$^{\circ}$ and fcc0$^{\circ}$ site, respectively. The calculated adsorption height from PBE+vdW$\rm^{surf}$ is in excellent agreement with the adsorption height as derived from LEED analysis (2.09 vs.\ $2.0 \pm 0.02$~{\AA})~\cite{wander1991}.
We also constructed a larger supercell of ($4 \times 4$) for Pt(111), and in this lower coverage case the adsorption energy is determined to be 2.18~eV from PBE+vdW$\rm^{surf}$. This lies within the uncertainty of calorimetry measurements in the limit of zero coverage (1.84--2.25~eV)~\cite{ihm2004}. 
% Further enlarging the size of the unit cell to ($5 \times 5$) only leads a 0.02~eV difference in the adsorption energy, with respect to the ($4 \times 4$) unit cell.

\subsection{Naphthalene on Ag(111)}
\paragraph{Experimental Data}
Naphthalene (Np) adsorbed on Ag(111) has been studied using LEED~\cite{2007Rockey,1988Frank,1975Firment}, NEXAFS~\cite{yannoulis1987}, 2PPE~\cite{2001Gaffney}, and TPD~\cite{2007Rockey,Ref2Huang}. Upon deposition at 90~K, the molecule is found to desorb completely at about 325~K for low initial coverages. This corresponds to an adsorption energy of $0.88 \pm 0.05$~eV assuming a pre-exponential of $10^{13}~$s$^{-1}~$~\cite{2007Rockey}. Recently, a factor of $10^{15.2}~$s$^{-1}~$ was suggested~\cite{PhysRevLett.115.086101} to be more accurate using the Campbell-Sellers method~\cite{Campbell2013, Campbell2013a} to estimate pre-exponential factors. The corresponding adsorption energy which we report in Tab.~\ref{tab:benzene} is 1.04~eV. For the molecule adsorbed in the monolayer a ($3 \times 3$) surface overlayer structure has been found from LEED~\cite{2007Rockey, 1988Frank}, whereas also different non-primitive overlayers have been observed at higher temperatures~\cite{1975Firment}. Using 2PPE, \citet{2001Gaffney} found that unoccupied states in the Np monolayer are mixed with image potential states of the interface. In mono- and multilayer arrangements, NEXAFS measurements found Np to adsorb flat on Ag(111)~\cite{yannoulis1987}.

\paragraph{Theoretical Data}
We modelled the Np/Ag(111) interface with PBE+vdW$^{\mathrm{surf}}$ as a ($4 \times 4$) unit cell with six substrate layers. In the optimal geometry of the molecule the phenyl rings are situated close to hollow sites or, equivalently, the central C--C bond is situated above a top site. The molecule adsorbs flat on the surface with no sign of hybridization or displacement of hydrogen atoms above or below the molecular plane. The average C--Ag vertical adsorption distance is 2.99~\AA{} at an adsorption energy of 1.22~eV.

\subsection{Naphthalene on Cu(111)}
\paragraph{Experimental Data}
Np on Cu(111) shows a variety of overlayer structures that are both commensurate and incommensurate, the latter giving rise to Moir\'e patterns~\cite{2010Yamada,Forker2014}. The overlayer structures are all non-primitive structures in surface area between (4x3),(5x3), and larger superstructures. All studies that performed TPD measurements~\cite{13Wang,2003Zhao,2001Lukas}, notwithstanding differences in absolute desorption temperatures and heating rates, identified a broad desorption feature believed to be associated with desorption from terraces and a sharp feature at lower temperatures associated with desorption from step edges~\cite{2001-2Lukas}. \citet{2001Lukas} report an adsorption energy of 0.81~eV for desorption from (111) terraces. Similar to the case of Np/Ag(111) we re-evaluate the desorption energy using a pre-exponential factor of $10^{15.2}$ as 1.07~eV.

\paragraph{Theoretical Data}
Equally as in the case of Np on Ag(111), we model Np/Cu(111) in a (4x4) unit cell with the molecule adsorbed flat on the surface and the conjugated phenyl rings situated above hollow sites. We find a PBE+vdW$^{\mathrm{surf}}$ adsorption energy of 1.41~eV and an average C--Cu vertical adsorption height of 2.73~\AA{}.

\subsection{Naphthalene on Pt(111)}
\paragraph{Experimental Data}
Np and other aromatic molecules on Pt(111) have been studied with LEED by different groups. An initial study proposed the overlayer structure of Np/Pt(111) to be a ($6 \times 6$) surface unit cell containing 4 molecules with alternating $90^{\circ}$ tilt angle with respect to each other~\cite{7Gland}. Later, Dahlgren and coworkers proposed a ($6 \times 3$) overlayer containing 2 molecules and satisfying a glide-plane symmetry~\cite{11Dahlgren}. In their model Np is adsorbed with its center at a top site and the 2 molecules are rotated by $60^{\circ}$ with respect to each other. The authors later also found that Np molecules on Pt(111) fully dehydrogenate above 200~K~\cite{121982Dahlgren}. The proposed overlayer structure was also later confirmed by STM and LEED~\cite{13Hallmark}. The adsorption energy of Np/Pt(111) has been measured by \citet{A2006Gottfried,11998Brown} using SCAM at 300~K. The authors identify heats of adsorption associated with adsorption at step edges and at terraces. Modelling the latter, they arrive at an adsorption energy of 3.11~eV for the zero-coverage limit. Higher molecule packing significantly decreases the adsorption energy. At a packing density of 0.59~ML corresponding to a (4x4) surface overlayer the adsorption energy is 2.19~eV (as reported in Tab.~\ref{tab:benzene}).

\paragraph{Theoretical Data}
A number of semi-empirical calculations have been performed by Gavezotti \emph{et al} for naphthalene on Pt(111)~\cite{1982Gavezotti,15Gavezotti}. The authors conclude on a bonding distance of 2.1~\AA{}. \citet{17Morin} report non-dispersion corrected DFT calculations of Np/Pt(111) using the PW91 functional. The authors predict a binding distance of 2.25~\AA{} and an adsorption energy of 0.53~eV for the experimental adsorption site, but find others to be more stable.

As before we model the Np/Pt(111) interface in a (4x4) overlayer using PBE+vdW$^{\mathrm{surf}}$. Contrary to Np adsorption at Cu(111) and Ag(111), we find significant distortion of the molecule upon adsorption in the most favorable adsorption site. As found in experiment, the most stable adsorption site is the molecule centered above a top site. We find the hydrogen atoms distorted away from the surface and out of the molecular plane. As a result the vertical adsorption height of Np at Pt(111) is 2.03~\AA{} for carbon atoms and 2.58~\AA{} for hydrogen atoms (see Tab.~\ref{tab:benzene}). We find the corresponding adsorption energy to be 2.92~eV which lies 0.73~eV above the experimental adsorption energy at the same coverage and 0.19~eV below the zero-coverage extrapolated adsorption energy.

\begin{table}
\caption{\label{tab:benzene} Adsorption energies (\textit{E}$\rm_{ad}$) and perpendicular heights (\textit{d}) for Bz and Np on (111) metal surfaces.
The values are in eV and \AA{}, respectively.
}
%\begin{ruledtabular}
\begin{tabular}{lcccc}
&\multicolumn{2}{c}{PBE+vdW$^{\rm surf}$} &\multicolumn{2}{c}{Exp.} \\
\cline{2-5} \noalign{\vskip 1pt} 
&\textit{E}$\rm_{ad}$ & \textit{d} & \textit{E}$\rm_{ad}$ & \textit{d} \\
\hline \noalign{\vskip 1pt} 
Bz/Cu(111) 		& 0.86 & 2.79 	& $0.69 \pm 0.04$~\cite{Liu2015} & -- \\
Bz/Ag(111) 		& 0.75 & 3.00 & $0.68 \pm 0.05$~\cite{Liu2015} & $3.04 \pm 0.02$~\cite{Liu2015} \\
Bz/Au(111) 		& 0.74 & 3.05 & $0.65 \pm 0.03$~\cite{Liu2015} &--\\
Bz/Pt(111) 		& 2.18 & 2.09  & 2.19~\cite{ihm2004} & $2.02 \pm 0.02$~\cite{wander1991}\\
Np/Cu(111) 	& 1.41 & 2.73  & 1.07$^a$ & -- \\
Np/Ag(111) 	& 1.22 & 2.99  & 1.04$^a$ & --  \\
Np/Pt(111) 	& 2.92& 2.03(C)/2.58(H) & 2.19$^b$ & -- \\ \hline \noalign{\vskip 3pt} 
\end{tabular}
$^a$: The reported adsorption energies have been recalculated based on desorption temperatures from TPD~\cite{2001Lukas,2007Rockey} and a pre-exponential factor of $10^{15.2}$ as estimated recently~\cite{PhysRevLett.115.086101}.\\
$^b$: measured heat of adsorption for a coverage of $\Theta=0.58$, which equals a ($4 \times 4$) surface unit cell~\cite{A2006Gottfried}.
\end{table}
%%%%%%%%%%%%%%%%%%%%%%%%%%%%%%%%%%%%%%%%%%%%%%%%

\section{Extended and Compacted Carbon Systems on Metal Surfaces}
\label{chapter-extended}

\subsection{DIP on Cu(111), Ag(111), and Au(111)}\label{DIP_exp}
\paragraph{Experimental Data}
The adsorption properties of Diindenoperylene (DIP, C$_{32}$H$_{16}$), a $\pi$-conjugated molecule, on noble metals has been extensively studied because of its excellent optoelectronic device performance and the ability to form exceptionally ordered films~\citep{DIP_ordered_1, DIP_ordered_2, DIP_gold}.
Also, the rather simple DIP structure and chemical composition, a planar hydrocarbon with no heteroatoms, make DIP/metal interfaces suitable to be used as model systems in the context of HIOSs.
The deposition of DIP on clean substrates, either metals or semiconductors, and the formation of an ordered monolayer has been observed in different studies using a variety of experimental techniques. The cleanliness of the metal substrate, the coverages, and the quality of the deposition has been studied using x-ray spectroscopy techniques such as XPS, NEXAFS and x-ray reflectivity techniques~\citep{DIP_xps, DIP_nexafs, DIP_ordered_2}.
The morphology and the electronic properties have been investigated with several methods, \textit{e.g.} TEM, STM, LEED and UPS spectroscopy~\citep{DIP_tem, DIP_copper, DIP_silver, DIP_gold, DIP_ordered_1}.
The adsorption of DIP on clean Au(111), Ag(111), and Cu(111) has been carefully monitored employing the experiments listed above and different possible interface structures are observed.
In general, the DIP surface density and arrangement can be influenced by the presence of step edges, terraces, or the substrate temperature  during growth~\citep{DIP_copper}.
DIP/Ag(111) has been investigated with low temperature STM and LEED, revealing different closed-packed monolayer configurations, namely a brick-wall and a herringbone arrangement~\citep{DIP_silver}.
Similarly to DIP/Ag(111), STM experiments found that DIP/Au(111) assumes a brick-wall configuration~\citep{DIP_gold}.
However, no experimental data are available for a thorough comparison of the cohesive energies.
Finally, recent NIXSW measurements extended the characterization of these systems, providing accurate average bonding distances of DIP on all three noble metals~\citep{Burker2013}. The measurements are listed in Tab.~\ref{tab-C60_DIP} and follow the trend: $d$(Cu)$< d$(Ag)$< d$(Au). 

\paragraph{Theoretical Data}
Electronic structure calculations for all the three metals were performed using a ($7 \times 7$) unit cell composed of three metal layers and one DIP molecule.
First we computed the adsorption energy curve by rigidly tuning the surface--molecule distance $d$ (see supplemental material).
As a second step, we relaxed the geometry with lowest $E_\mathrm{ad}$ for each system.
The average bonding distances obtained from our simulations show a remarkably good agreement with the experimental data, see Tab.~\ref{tab-C60_DIP}, with a discrepancy of less than 0.1\,\r{A} for all the three systems.
Moreover, in accordance with the bonding distances, the adsorption energies show the trend: 
$|E_{\rm ad}{\rm (Cu)}|>|E_{\rm ad}{\rm (Ag)}|>|E_{\rm ad}{\rm (Au)}|$.
In the case of DIP/Ag(111), in addition to the structures considered above, we take into account the two densely-packed and well-ordered configurations of the monolayer: the brick-wall and the herringbone arrangements.
Remarkably, the bonding distance obtained from the relaxed geometries is $d=2.99$\,\r{A} in both cases.
This equilibrium distance is in  almost perfect agreement with the XSW measurement, even improving the result reported in Tab.~\ref{tab-C60_DIP}.
Furthermore, the relaxed structure obtained using the brick-wall arrangement for DIP/Au(111) yields a binding distance of 3.15\,\r{A},  also in better agreement with the experiment than the ($7 \times 7$) structure.

\subsection{C$_{60}$ on Au(111) and Ag(111)}
\begin{figure}
\includegraphics[width=\columnwidth]{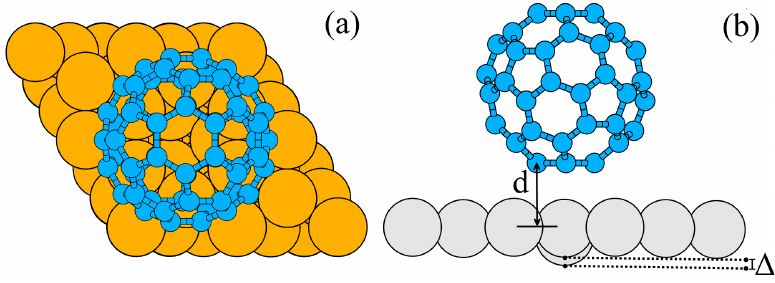}
\caption{\label{fig-C60} Adsorption geometry of C$_{60}$ on Au(111) and Ag(111). (a) Top view of the unit cell. (b) Side view with one metal layer. $d$ indicates the surface-molecule distance and $\Delta$ is the buckling amplitude of the first metal layer.}
\end{figure}

\paragraph{Experimental Data}\label{C60_exp}
The fullerene C$_{60}$ molecule has been intensively studied since its discovery, for its interesting properties such as superconductivity~\citep{c60_superconductor} or metal-insulator transition~\citep{c60_transition}.
In the same spirit, the deposition of thin films of C$_{60}$ on noble metal surfaces opens up numerous possible applications, \textit{e.g.} lubrication and molecular switching.
These properties, combined with the ability of C$_{60}$ to form ordered monolayers on surfaces, motivated a large number of experiments performed with several different techniques, \textit{e.g.} UHV-STM, LEED, Auger electron spectroscopy (AES).
It was found that C$_{60}$ adsorbed on Ag(111) and Au(111) is adsorbed in a well-ordered closed-packed monolayer and displays a commensurate $(2\sqrt{3} \times 2\sqrt{3})R30^{\circ}$ unit cell~\cite{c60_stm1, c60_stm3}.
C$_{60}$ adsorbs in different sites and it is also possible to manipulate the facet of the molecule exposed to the surface, \textit{e.g.} two distinct orientations are found by tuning the level of potassium doping~\cite{c60_xpd}.
A UHV-STM and LEED study concluded that the molecule adsorbs, on both Ag and Au(111), preferably on top sites with a pentacene ring facing  down~\cite{c60_stm2}. On the other hand, for C$_{60}$/Ag(111), a STM experiment revealed that the adsorption took place on hollow sites, with a hexagonal ring facing the surface~\cite{c60_stm5}.
Moreover, a mix of hexagonal face-down and C-C bond down was found in an x-ray photoelectron diffraction study~\cite{c60_xpd}.
Within the different possible monolayer configurations, recent LEED experiments of C$_{60}$ on Ag(111) confirm the $(2\sqrt{3} \times 2\sqrt{3})R30^{\circ}$ unit cell and suggest that the most stable configuration is with the molecule situated above a vacancy site of the metal surface.
The C$_{60}$ molecule is adsorbed on a vacant-top site with a hexagon face-down orientation~\cite{c60_leed}.
During the adsorption, the surface relaxes and the silver atoms close to the C$_{60}$ are slightly compressed into the surface. This is indicated with the distance measure $\Delta$ in Fig.~\ref{fig-C60} panel b.
This interlayer buckling amplitude and the distance between the molecule and the topmost layer are computed by fitting LEED measurements.
Similarly, STM images for C$_{60}$ on Au(111) illustrate the presence of a vacancy, confirming that the molecule is adsorbed above the vacancy with the hexagon face-down configuration~\cite{c60_stm4}.
To the best of our knowledge, no experimental data are available for the energetics of these two systems.

\paragraph{Theoretical Data}\label{C60_th}
In order to reproduce the experimentally observed structure, we considered, for both silver and gold, a $(2\sqrt{3} \times 2\sqrt{3})R30^{\circ}$ unit cell with six metal layers.
This unit cell contains one C$_{60}$ molecule, as shown in Fig.~\ref{fig-C60}a, placed in the center of the cell and adsorbed in correspondence of a top site.
Further, this particular top atom is removed creating a single vacancy site on the topmost metal layer.
We performed a relaxation involving all the six metal layers in order to capture also the intra-layer rearrangement and obtain a better comparison between our simulations and the experimental findings.
Notably, different molecular orientations have been taken into consideration as a starting point of our simulations.
However, during the relaxation procedure, the molecule rearranges to a hexagonal face-down orientation, confirming that the latter structure represents the most stable geometry in the presence of a vacancy.
The binding energies for both systems are computed taking into account the formation energy of the vacancy (see supplemental material) and are reported in Tab.~\ref{tab-C60_DIP}.
We find that C$_{60}$ binds stronger to Au(111) than to Ag(111).

The predicted binding distances, for both Ag(111) and Au(111), are in excellent agreement with the experimental data, as reported in Tab.~\ref{tab-C60_DIP}. 
Considering the buckling amplitude $\Delta$,
the adsorption of C$_{60}$ on Ag(111) produces a $\Delta$ of 0.02\,\r{A} and 0.03\,\r{A} for the first and second layer respectively~\cite{c60_leed}.
From our simulations we find $\Delta=0.015$\,\r{A} for the first layer and a larger $\Delta=0.03$\,\r{A} for the second, confirming the experimental trend.
In the case of C$_{60}$ on Au(111), the experimental measurements indicate a decrease in the amplitude from 0.05\,\r{A} to 0.02\,\r{A} within the first two metal layers~\cite{c60_au111}.
The $\Delta$ obtained from the theoretical calculations reproduce the same trend, with slightly larger values: 0.08\,\r{A} and 0.045\,\r{A} respectively.

\begin{table}
\caption{\label{tab-C60_DIP} Adsorption energies (in eV) and perpendicular heights (in \AA{}) for DIP and C$_{60}$ on (111) metal surfaces.
}
%\begin{ruledtabular}
\begin{tabular}{lcccc}
&\multicolumn{2}{c}{PBE+vdW$^{\rm surf}$}  &\multicolumn{2}{c}{Exp.} \\
\cline{2-5} \noalign{\vskip 1pt} 
&\textit{E}$\rm_{ad}$ & \textit{d} & \textit{E}$\rm_{ad}$ & \textit{d} \\
\hline \noalign{\vskip 1pt} 
DIP/Cu(111) 			&4.74 &2.59 & -- & $2.51 \pm 0.03$~\citep{Burker2013}\\
DIP/Ag(111) 			&3.55 &2.94 & -- & $3.01 \pm 0.04$~\citep{Burker2013}\\
DIP/Au(111) 			&2.53 &3.22 & -- & $3.17 \pm 0.03$~\citep{Burker2013}\\
C$_{60}$/Ag(111) 		&2.82 &1.99 & -- & $2.01 \pm 0.1$~\cite{c60_leed}\\
C$_{60}$/Au(111) 		&3.36 &1.81 & -- & $1.8 \pm 0.1$~\cite{c60_au111}\\
\hline
\end{tabular}
\end{table}
%%%%%%%%%%%%%%%%%%%%%%%%%%%%%%%%%%%%%%%%%%%%%%%%

\section{Sulfur-containing Systems on Metal Surfaces}
\label{chapter-sulfur}

\subsection{Thiophene on Au(111)}
\paragraph{Experimental Data}
Thiophene (Thp) is one of the smallest heteroaromatic molecules for which metal-surface adsorption has been studied.
Thp has been adsorbed on Au(111) surfaces from vacuum~\cite{2003Nambu,2001Liu} and from solution~\cite{2002Matsuura,Sako2005}, whereas only vacuum adsorption leads to successful adsorption of pristine Thp molecules.
Adsorption from ethanol solution leads to decomposition of Thp molecules as evidenced by infrared absorption spectroscopy, XPS, and NEXAFS measurements~\cite{Sako2005}

Using TPD, XPS, and NEXAFS experiments \citet{2001Liu} and \citet{2003Nambu} have studied Thp adsorbed on Au(111).
Thp adsorbs below 100~K and desorbs fully already around 330~K.
Both above mentioned TPD studies find slightly different desorption temperatures, due to different experimental conditions on heating rate (2~K/s vs. 3~K/s) and adsorption temperature.
However, both studies find a desorption peak at low coverage at 215~K (255~K) that is associated with a flat-lying adsorbate structure and related to average adsorption energies of 0.60~eV~\cite{2003Nambu} (0.70~eV~\cite{2001Liu}).
STM experiments support the assessment that this phase corresponds to flat-lying geometries~\cite{1996Dishner}.

A second TPD peak around 180~K appears at higher coverages and saturates at a nominal coverage of 2.3~ML.
This signature is associated with a compressed monolayer structure of tilted Thp molecules and an effective adsorption energy of 0.47~eV by both studies.
At higher coverage a third signature at 140-150~K is found that does not saturate with coverage, consistent with desorption from a physisorbed multilayer.

The geometry change from a flat-lying to a tilted compressed monolayer is corroborated by XPS and NEXAFS experiments.
With increasing coverage above 1~ML the flat-lying phase gradually transforms to a tilted compressed phase with a tilt angle of $55^{\circ}$ with respect to the surface parallel~\cite{2003Nambu}.
The LEED pattern associated with this phase is a $(\sqrt{3}\times\sqrt{3})R30^{\circ}$. Unfortunately no experimental adsorption heights have been reported for Thp/Au(111).
STM experiments suggest that preferential adsorption occurs at the hcp hollow site~\cite{2003Gui-Jin}, however due to paired-row structures at high coverage,  Thp can coexist at different adsorption sites~\cite{2002Noh}.
A notable feature of Thp on Au surfaces is an X-ray induced polymerization reaction to polythiophene films~\cite{Salaneck1987}.

\paragraph{Theoretical Data}
The Thp/Au(111) interface has been modelled using pure semi-local DFT and variants of the DFT-D method~\cite{Tonigold2010}.
Using the PBE functional~\cite{Perdew1996} to model the flat-lying Thp molecule \citet{Tonigold2010} find a vanishingly small adsorption energy at a surface-sulfur distance of 3.4~\AA{}.
Incorporating dispersion interactions in the form of two variants of DFT-D, the authors find adsorption energies of 1.24~ and 1.73~eV and surface-sulfur distances of 2.75 and 2.88~{\AA}, respectively.
In these geometries, the sulfur atom lies closer to the surface than the rest of the aromatic ring.

Using DFT+vdW$^{\mathrm{surf}}$, we have modelled a single Thp molecule at a Au(111) surface in a ($3 \times 3$) unit cell with the molecule lying flat on the surface, as shown in Fig.~\ref{fig-bz-thioph}(b).
We find the most stable adsorption geometry for the sulfur atom situated at a top site and the aromatic ring centered closely above a hcp hollow site.
This is the most stable adsorption geometry on Cu(111), Ag(111), and Au(111).
The corresponding adsorption height of Thp/Au(111) as measured from the sulfur atom is 2.95~\AA{} (see Tab.~\ref{tab-thiophene}) with an adsorption energy of 0.77~eV.
The binding energy as described by DFT+vdW$^{\mathrm{surf}}$ is significantly reduced compared to the results of Tonigold \emph{et al.}, as is expected from the inclusion of substrate screening effects. Analysing the desorption temperatures of experiment using the Redhead equation~\cite{Redhead1962} and preexponential factors~\cite{Campbell2012,Campbell2013} in the range of $10^{12}$ to $10^{15}$~s$^{-1}$ we find adsorption energies of 0.53 to 0.66~eV~\cite{2003Nambu}  (0.62 to 0.78~eV~\cite{2001Liu}), which deviate from the DFT+vdW$^{\mathrm{surf}}$ by 0.01~eV for the experimental upper limit.

\subsection{Thiophene on Ag(111)}
\paragraph{Experimental Data}
TPD measurements of Thp on Ag(111) from 2 different studies~\cite{1991Baumgartner,2000Vaterlein} identified three desorption peaks corresponding to different overlayer structures: initial desorption at 128~K (140~K) is associated with Thp molecules in the multilayer, a desorption peak at 148~K (162~K) is associated with Thp molecules directly adsorbed on Ag(111) in a densely packed tilted arrangement, and desorption at 190~K (204~K) is believed to arise from molecules that are adsorbed in a flat arrangement on the surface.
Discrepancies between the absolute desorption temperatures of the two studies are believed to arise from differing approaches to temperature measurement in the experimental setup~\cite{2000Vaterlein}.
C K-edge NEXAFS measurements support the assessment that the high-temperature feature corresponds to molecules adsorbed in an almost flat arrangement~\cite{2000Vaterlein}. \citet{1996Chen} identified three monolayer structures of Thp/Ag(111) from STM experiments upon deposition of 40, 90, and 150ML, respectively: a c($2\sqrt{3}\times4$)rect structure, a  ($2\sqrt{7}\times2\sqrt{7}$) structure, and a herringbone structure. These overlayer structures show largely different packing densities.
The authors infer from the area per molecule that the c($2\sqrt{3}\times4$)rect structure corresponds to a flat-lying molecular geometry and the others to more strongly tilted adsorbates. 
On the basis of the measured desorption temperatures of the above mentioned studies we estimate the adsorption energy for the flat-lying phase using the Redhead equation to be between 0.48 and 0.64~eV for preexponential factors between $10^{12}~s^{-1}$ to $10^{15}~s^{-1}$.

\paragraph{Theoretical Data}
We have studied Thp on Ag(111) in a ($3 \times 3$) surface unit cell using DFT+vdW$^{\mathrm{surf}}$. As in the case of Thp/Au(111), we find as optimal geometry a flat-lying molecule with a S--Ag distance of 3.17~\AA{} and an adsorption energy of 0.72~eV, which overestimates the experimental regime by 0.08~eV.

\subsection{Thiophene on Cu(111)}
\paragraph{Experimental Data}
Milligan and coworkers performed a combined NIXSW, NEXAFS, and TPD study of Thp on Cu(111) and identified two different monolayer phases~\cite{1998Milligan,2001Milligan}.
At low coverage Thp adsorbs in an almost flat-lying geometry, ranging from 12 to $25 \pm 5^{\circ}$ for 0.03 to 0.1~ML coverage with respect to the surface and a Cu--S bond distance of $2.62 \pm 0.03$~\AA.
The corresponding desorption maximum lies at a temperature of 234~K.
At higher coverages a second coexisting phase with higher packing density forms.
The Thp molecules are tilted more strongly with a surface angle of $44 \pm 6^{\circ}$ and a S--Cu distance of $2.83 \pm 0.03$~\AA.
The corresponding desorption peak is found at 173~K.
The authors further find that the sulfur atom of Thp predominantly adsorbs at atop sites~\cite{1998Milligan}.
A number of other studies support the finding of a flat-lying adsorbate phase~\cite{2002Rousseau,1981Richardson}.
With the use of the Redhead equation~\cite{Redhead1962} and the experimental heating rate of 0.5~K/s we can estimate the experimental binding energy of the flat-lying phase of Thp from the data of \citet{2001Milligan}.
Assuming a preexponential factor in the range of $10^{12}$ to $10^{15}~s^{-1}$, the adsorption energy is 0.61 to 0.75~eV.

\paragraph{Theoretical Data}
On Cu(111), Thp in a ($3 \times 3$) overlayer unit cell also adsorbs preferentially with sulfur situated at an atop--site. The corresponding S--Cu distance is 2.78~\AA{}, which is in good agreement with experiment. We find the molecule adsorbed with a tilt angle of $6^{\circ}$, which is significantly less than the experimentally found angle of $12 \pm 5^{\circ}$ for 0.03~ML. The reason for this discrepancy could be an increased tilt angle at finite temperature due to anharmonicity of the adsorbate-substrate bond. The calculated adsorption energy of 0.82~eV overestimates the experimental range by 0.07~eV.

\begin{table}
\caption{\label{tab-thiophene} Adsorption energies (in eV) and average perpendicular heights of Carbon (C) and Sulfur (S) atoms (in \AA{}) for Thp on Au, Ag, and Cu (111) surfaces.
}
%\begin{ruledtabular}
\begin{tabular}{lcccc}
&\multicolumn{2}{c}{PBE+vdW$^{\rm surf}$} &\multicolumn{2}{c}{Exp.} \\
			 \cline{2-5} \noalign{\vskip 1pt} 
&\textit{E}$\rm_{ad}$ & \textit{d} & \textit{E}$\rm_{ad}$ & \textit{d} \\
\hline \noalign{\vskip 1pt} 

	Thp/Au(111) 	&0.77 	&2.95(S) -- 3.01(C)  	& 0.53 -- 0.78$^a$ 	& -- 		\\
	Thp/Ag(111) 	&0.72 	&3.17(S) -- 3.19(C) 	&0.48 -- 0.64$^a$ 		& -- 		\\
	Thp/Cu(111) 	&0.82 	&2.78(S) -- 2.9(C) 	& 0.61 -- 0.75$^a$ 	& $2.62 \pm 0.05$ \\
\hline \noalign{\vskip 2pt}
\end{tabular}
$^a$: Experimental adsorption energies estimated from desorption temperatures of Refs.  \onlinecite{2001Milligan,1996Chen,2001Liu,2003Nambu} using Redhead equation~\cite{Redhead1962} with preexponential factors of $10^{12}$ to $10^{15}~s^{-1}~$~\cite{Campbell2012,Campbell2013}.
\end{table}

%%%%%%%%%%%%%%%%%%%%%%%%%%%%%%%%%%%%%%%%%%%%%%%%

\section{Oxygen-containing Systems on Metal Surfaces}
\label{chapter-oxygen}

\begin{table*}
\centering
\caption{Comparison of experimental and theoretical results for the adsorption geometry of PTCDA on Ag(111), Ag(100), and Ag(110). We use $d_{\rm Th/Exp}$ to denote the vertical adsorption heights, given in \AA{}, of the specific atoms obtained from PBE+vdW$^{\mathrm{surf}}$ calculations and NIXSW studies. The specification of the atoms can be seen in Fig.~\ref{Fig_PTCDA_Au111Struct}(c). The C backbone distortion is given as  
$\Delta \mathrm{C} = d(\mathrm{C_{peryl}})-d(\mathrm{C_{func}})$ and the O difference as 
$\Delta \mathrm{O} = d(\mathrm{O_{anhyd}})-d(\mathrm{O_{carb}})$. Experimental results can be found in 
Refs.~\onlinecite{Hauschild2010,Bauer:Mercurio:etalPRB2012,Mercurio:Bauer:etalPRB2013}. We cite here the results 
given in  Refs.~\onlinecite{Hauschild2010,Bauer:Mercurio:etalPRB2012}. In addition, we show the adsorption energies $E_{\mathrm{ad}}$ in eV for each system with the PBE+vdW$^{\mathrm{surf}}$ method.}
\setlength{\tabcolsep}{10pt}
\begin{tabular}{ccccccc}
%\hline
 & \multicolumn{2}{c}{Ag(111)} & \multicolumn{2}{c}{Ag(100)} & \multicolumn{2}{c}{Ag(110)} \\
\cline{2-7} \noalign{\vskip 2pt} 
& $ d_{\rm Th} $ & $d_{\rm Exp}$~\cite{Hauschild2010} & $ d_{\rm Th} $ & $d_{\rm Exp}$~\cite{Bauer:Mercurio:etalPRB2012} & $ d_{\rm Th} $ & $d_{\rm Exp}$~\cite{Bauer:Mercurio:etalPRB2012}\\
\hline \noalign{\vskip 1pt} 
C total              & 2.80 & $2.86 \pm 0.01$ & 2.75 & $2.81 \pm 0.02$ & 2.54 & $2.56 \pm 0.01$ \\
C$_{\mathrm{peryl}}$ & 2.80 & --              & 2.76 & $2.84 \pm 0.02$ & 2.56 & $2.58 \pm 0.01$  \\
C$_{\mathrm{func}}$  & 2.78 & --              & 2.67 & $2.73 \pm 0.01$ & 2.43 & $2.45 \pm 0.11$  \\
$\Delta \mathrm{C}$  & 0.02 & --              & 0.09 & $0.11 \pm 0.02$ & 0.13 & $0.13 \pm 0.11$  \\
O total 	     & 2.73 & $2.86 \pm 0.02$ & 2.59 & $2.64 \pm 0.02$ & 2.33 & $2.33 \pm 0.03$  \\
O$_{\mathrm{carb}}$  & 2.68 & $2.66 \pm 0.03$ & 2.54 & $2.53 \pm 0.02$ & 2.29 & $2.30 \pm 0.04$  \\
O$_{\mathrm{anhyd}}$ & 2.83 & $2.98 \pm 0.08$ & 2.69 & $2.78 \pm 0.02$ & 2.39 & $2.38 \pm 0.03$   \\
$\Delta \mathrm{O}$  & 0.15 & $0.32 \pm 0.09$ & 0.15 & $0.25 \pm 0.02$ & 0.10 & $0.08 \pm 0.05$   \\
\hline \noalign{\vskip 1pt} 
 $E_{\mathrm{ad}}$   & 2.86 & --	& 2.93 & 	--	 & 3.39 &  	--	\\
\hline 
\end{tabular}
\label{Table_PTCDAAgSurfaces}
\end{table*}
%%%%%%%%

\subsection{PTCDA on Ag(111), Ag(100), and Ag(110)}
\paragraph{Experimental Data}
Up to this point we have mostly addressed the performance of the DFT+vdW$^{\mathrm{surf}}$ method in the adsorption of atoms and molecules on close-packed (111) transition-metal surfaces, but we are also interested in the performance and sensitivity of the DFT+vdW$^{\mathrm{surf}}$ method when the adsorption occurs on non-close-packed surfaces. We address this aspect by reviewing a comparative analysis of the adsorption of PTCDA on a surface with different orientations: PTCDA on Ag(111), Ag(100), and Ag(110) (see Ref.~\onlinecite{Ruiz:Liu:Tkatchenko2016}). PTCDA is a chemical compound formed by an aromatic perylene core (C$_{\mathrm{peryl}}$) terminated with two anhydride functional groups, each of them containing two carbon atoms (C$_{\mathrm{func}}$), two carboxylic oxygens (O$_{\mathrm{carb}}$) and one anhydride oxygen (O$_{\mathrm{anhyd}}$) \cite{Bauer:Mercurio:etalPRB2012}. The adsorption geometries of these systems have been investigated using the NIXSW technique.~\cite{Hauschild2010,Bauer:Mercurio:etalPRB2012,Mercurio:Bauer:etalPRB2013} A novel feature in the studies including PTCDA on Ag(100) and Ag(110) is their higher chemical resolution resulting in the 
extraction of the adsorption positions of each of the chemically inequivalent atoms in PTCDA. 

PTCDA forms a commensurate monolayer structure on silver surfaces. On Ag(111), it forms a herringbone structure with two molecules per unit cell in non-equivalent adsorption configurations.~\cite{Gloeckler:Seidel:etalSS1998,Kraft:Temirov:etalPRB2006} Both molecules are adsorbed above a bridge site, molecule A is aligned with the substrate in the $[10\bar{1}]$ direction with its carboxylic oxygen atoms on top position and the anhydride oxygen atoms located on bridge sites. Molecule B on the other hand is rotated with respect to the $[01\bar{1}]$ direction, with most atoms in its functional groups located closely to adsorption bridge positions.~\cite{Kraft:Temirov:etalPRB2006,Bauer:Mercurio:etalPRB2012} On Ag(100), a T-shape arrangement with two adsorbed molecules per unit cell can be observed.~\cite{Ikonomov:Bauer:SokolowskiSS2008} Both molecules are aligned with the $[110]$  direction of the substrate with the center of each molecule adsorbed on top position. Finally, in the case of Ag(110), PTCDA forms a brick-wall adsorption pattern with one molecule adsorbed per surface unit cell.~\cite{Gloeckler:Seidel:etalSS1998} The long axis of the molecule is located parallel to the $[001]$ direction, while the center of the molecule is located on the bridge site between the close-packed atomic rows parallel to the $[\bar{1}10]$ direction.~\cite{Bohringer:Schneider:etalSS1998}

\paragraph{Theoretical Data}
In the following we discuss results of Refs.~\onlinecite{Ruiz:Liu:Tkatchenko2016} and \onlinecite{VRuizPhDThesis}. Tab.~\ref{Table_PTCDAAgSurfaces} shows that the PBE+vdW$^{\mathrm{surf}}$ results for the vertical adsorption distance agree very well with experimental results. With the exception of the anhydride oxygen in Ag(111), the calculated distances for all atoms that form the molecule lie within 0.1~\AA{} of the experimental results for all three surfaces. These results also reveal that our calculations reproduce the experimental trends observed in the sequence of Ag(111), Ag(100), and 
Ag(110).~\cite{Bauer:Mercurio:etalPRB2012,Mercurio:Bauer:etalPRB2013} The overall vertical adsorption height given by the calculations, taken as an average over all carbon atoms (see $d_{\mathrm{Th}}$ for `C total' in Tab.~\ref{Table_PTCDAAgSurfaces}), decreases in the sequence  of Ag(111), Ag(100), and Ag(110) by 0.26~\AA{} in comparison to the decrease of 0.30~\AA{} obtained in experiments. The calculations reproduce the transition from a \emph{saddle}-like adsorption geometry of PTCDA on Ag(111) to the \emph{arch}-like adsorption geometry that can be found in the more open surfaces according to experiments  (see Ref.~\onlinecite{Ruiz:Liu:Tkatchenko2016}). Finally, for the above mentioned sequence, we find an increase in the C backbone distortion and a decrease in the O difference ($\Delta \mathrm{C}$ and $\Delta \mathrm{O}$ defined in Tab.~\ref{Table_PTCDAAgSurfaces}). For $\Delta \mathrm{C}$, the calculations yield 0.02, 0.09, and 0.13~\AA{} for Ag(111), Ag(100), and Ag(110), respectively, values which are in excellent agreement with experiments.~\cite{Bauer:Mercurio:etalPRB2012,Mercurio:Bauer:etalPRB2013} In the case of Ag(111), the C backbone distortion has not been determined experimentally,~\cite{Hauschild2010} but the \emph{saddle}-like adsorption geometry suggests a minimum distortion of the C backbone~\cite{Hauschild2010,Bauer:Mercurio:etalPRB2012}, which we observe in our 
calculations as well. The C backbone distortion in Ag(100) and Ag(110) is then remarkably well reproduced by the calculations.  

With respect to the oxygen difference ($\Delta \mathrm{O}$), the resulting values are 0.15~\AA{} for Ag(111) and Ag(100), and 0.10~\AA{} for Ag(110). These values reproduce the decrease in the sequence observed by experiments but underestimate the difference by 0.17~\AA{} in Ag(111) and 0.10~\AA{} in Ag(100). This underestimation lies in the fact that the adsorption distances for the anhydride oxygen obtained with the calculations are also underestimated in the cases of  Ag(111) and Ag(100). On the other hand, the calculated distance of the anhydride oxygen to the other oxygen atoms agrees very well with the experimental result of $0.08 \pm 0.05$~\AA{} for PTCDA on Ag(110)~\cite{Bauer:Mercurio:etalPRB2012}.

We summarize calculated PBE+vdW$^{\mathrm{surf}}$ adsorption energies in Tab.~\ref{Table_PTCDAAgSurfaces}. The binding strength increases in the sequence $2.86$, $2.93$, and $3.39$ eV for Ag(111), Ag(100), and Ag(110), respectively. The vdW interactions are essential in these systems as they yield the largest contribution to the adsorption energy (see Ref.~\onlinecite{Ruiz:Liu:Tkatchenko2016} for details). We note that $E_{\mathrm{ad}}$ is calculated with respect to the PTCDA monolayer. The binding strength becomes even larger when calculated with respect to the molecule in gas phase due to the additional contribution of the monolayer formation energy. The accuracy of these results confirm the sensitivity to surface termination that the DFT+vdW$^{\mathrm{surf}}$ scheme correctly reproduces.

\subsection{PTCDA on Au(111)}
\label{PTCDAonAu111}

\paragraph{Experimental Data}
In the case of the Au(111) surface, PTCDA does not form a commensurate monolayer but rather exhibits a situation very close to a point--on--line correspondence with the $(22 \times \sqrt{3})$ reconstructed surface~\cite{Fenter:Schreiber:etalPRB1997,Kilian:Umbach:SokolowskiSS2006,Mannsfeld:Toerker:etalOE2001,Schmitz-Huebsch:Fritz:etalPRB1997}. \citet{Kilian:Umbach:SokolowskiSS2006} reported an adsorbate structure at equilibrium conditions (grown at high substrate temperatures and small deposition rates) that suggests an optimal point--on--line relation on each of the three reconstruction domains of the substrate, which results in azimuthal domain boundaries (with an angular misfit of around $2.5^\circ$) in the PTCDA monolayer. The adsorbate structure consists of a rectangular unit cell with an area of approximately 232~\AA$^2$ and two PTCDA molecules per surface unit cell. PTCDA is physisorbed on Au(111) and its bonding interaction is governed mainly by vdW forces.~\cite{Henze:Bauer:etalSS2007,Tautz2007,Wagner2012,Ziroff:Gold:etalSS2009} \citet{Wagner2012} studied the system based on single molecule manipulation experiments. By combining STM and frequency-modulated AFM, they reported an adsorption energy of about $2.5$ eV per molecule of PTCDA and an adsorption distance of approximately 3.25~\AA{}.
On the other hand, TPD experiments report an adsorption energy of approximately $-1.93 \pm 0.04$~eV per molecule in the low coverage limit~\cite{Stremlau:PTCDAAu111}.
The case of PTCDA on Au(111) has also been measured using the NIXSW technique~\citep{Henze:Bauer:etalSS2007,Hauschild2010} where they found an adsorption distance of $3.34 \pm 0.02$~\AA{} for the PTCDA monolayer.~\citet{Henze:Bauer:etalSS2007} reported, however, an adsorption height that corresponds most probably to the square phase of PTCDA on Au(111)~\cite{Mannsfeld:Toerker:etalOE2001,Schmitz-Huebsch:Fritz:etalPRB1997}, which does not conform the (majority) herringbone type phases observed by LEED~\cite{Fenter:Schreiber:etalPRB1997,Mannsfeld:Toerker:etalOE2001}. Accounting for an estimated outward relaxation of the topmost metal layer by 3\%, the authors report an adsorption height of $3.27 \pm 0.02$~\AA.

\begin{figure}
\begin{center}
\includegraphics[width=0.5\textwidth]{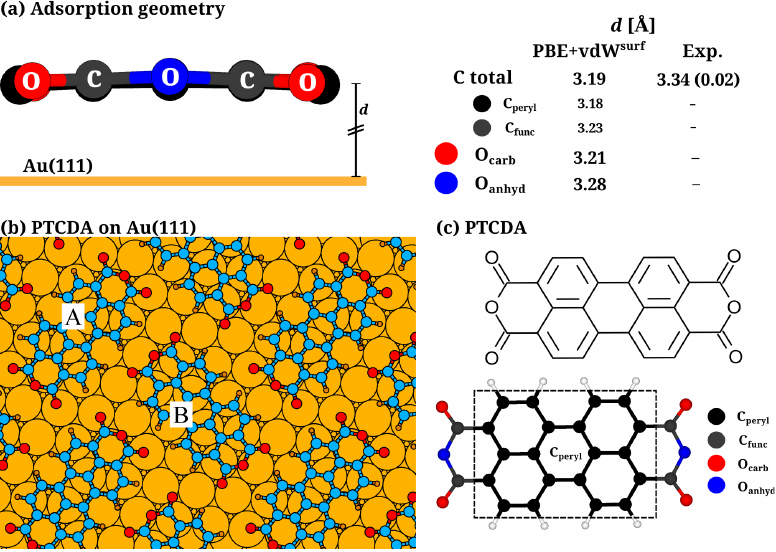}
\caption[(a) Structure of PTCDA adsorbed on Au(111) where the equilibrium distances $d$ for each chemically inequivalent atom calculated with the PBE+vdW$^{\mathrm{surf}}$ method are displayed. (b) Top view of the relaxed structure of PTCDA on Au(111). (c) Chemical structure of PTCDA.]{(a) Structure of PTCDA adsorbed on Au(111) where the equilibrium distances $d$ for each chemically inequivalent atom calculated with the PBE+vdW$^{\mathrm{surf}}$ method and measured by experiment\citep{Henze:Bauer:etalSS2007} are displayed. (b) Top view of the relaxed structure of PTCDA on Au(111). Both inequivalent molecules of the structure are labeled as A and B.  (c) Chemical structure of PTCDA. Images of the structures were produced using the visualization software VESTA \cite{VESTA}.}\label{Fig_PTCDA_Au111Struct}
 \end{center}
\end{figure}

\paragraph{Theoretical Data}
Tab.~\ref{Table_PTCDAAu111} shows the average vertical distance of each species in the PTCDA molecule.
Experimental results~\cite{Henze:Bauer:etalSS2007} are also shown for comparison.  Fig.~\ref{Fig_PTCDA_Au111Struct} depicts the structure of the monolayer after relaxation showing the position of each of the two inequivalent molecules in the unit cell. We model the system using a $\bigl(\begin{smallmatrix} 6 & 1 \\ -3 & 5 \end{smallmatrix} \bigr)$ surface unit cell which has an area of 247~\AA$^2$ (See supplemental material).

\begin{table}
\centering
\caption{Experimental and theoretical results for the adsorption geometry of PTCDA on Au(111). $d_{\rm Th/Exp}$ denotes the averaged vertical adsorption heights (in \AA{}) obtained with PBE+vdW$^{\mathrm{surf}}$ calculations and NIXSW studies. The atom nomenclature is given in Fig.~\ref{Fig_PTCDA_Au111Struct}(c). Experimental results for the adsorption distance can be found in Ref.~\onlinecite{Henze:Bauer:etalSS2007}. An estimated experimental adsorption height, which takes into account an outward relaxation of the topmost metal layer by 3\%, yields an adsorption height of the C backbone of 3.27~\AA.}  
\setlength{\tabcolsep}{12pt}
\begin{tabular}{c|cccccc}
& $ d_{\rm Th} $ & $d_{\rm Exp}$~\cite{Henze:Bauer:etalSS2007}  \\
\hline
C total              & 3.19 & $3.34 \pm 0.02$  \\
C$_{\mathrm{peryl}}$ & 3.18 & --               \\
C$_{\mathrm{func}}$  & 3.23  & --               \\
$\Delta \mathrm{C}$  & -0.05  & --               \\
O total 	     & 3.23 & --   \\
O$_{\mathrm{carb}}$  & 3.21 & -- \\
O$_{\mathrm{anhyd}}$ & 3.28 & -- \\
$\Delta \mathrm{O}$  & 0.07 & -- \\
\hline
\end{tabular}
\label{Table_PTCDAAu111}
\end{table}
%%%%%%%%

%%%%

Calculations with the PBE+vdW$^{\mathrm{surf}}$ method result in an adsorption height of 3.19~\AA{} for the C backbone, underestimating the experimental result~\cite{Henze:Bauer:etalSS2007} of 3.34~\AA{} by approximately 0.15~\AA{}. The positions of the O atoms were not measured in experiment due to an overlap of Au Auger lines with the O 1s core level. The results suggest a minor distortion of the C backbone as shown by $\Delta \mathrm{C}=-0.05$~\AA{}. The negative sign in $\Delta \mathrm{C}$ indicates that the C atoms belonging to the functional groups are located at a higher position than the C atoms of the perylene core. This fact is also reflected in the average position of the O atoms which is around 0.4~\AA{} higher than the average C backbone position. The anhydride O atoms are located around 0.09~\AA{} higher than the C backbone. Overall, the large adsorption height of the monolayer confirms a relatively weak interaction of the molecule with the surface in comparison to adsorption on Ag surfaces.

The discrepancy of around 0.15~\AA{} between the PBE+vdW$^{\mathrm{surf}}$ results and experiment can be attributed to several factors related to both approaches. First of all,  the area of the surface unit cell here studied is 247~\AA$^2$, which is larger than the area of the experimental one by more than 6\%. On the other hand, \citet{Henze:Bauer:etalSS2007} reported an adsorption height that corresponds most probably to the square phase of PTCDA on Au(111)~\cite{Mannsfeld:Toerker:etalOE2001,Schmitz-Huebsch:Fritz:etalPRB1997}, which does not conform the (majority) herringbone type phases observed by LEED~\cite{Fenter:Schreiber:etalPRB1997,Mannsfeld:Toerker:etalOE2001}. Finally, neither theory nor experiments take initially into consideration the surface relaxation in the determination of the adsorption height of the C backbone. Taking the estimated experimental~\cite{Henze:Bauer:etalSS2007} adsorption height of $3.27 \pm 0.02$~\AA{} as reference, which takes into account an estimated outward relaxation of the topmost metal layer by 3\%, the difference between theory and experiment is reduced to less than 0.1~\AA{}. Although the correct superstructure of the monolayer including the domain boundaries cannot be achieved by any state-of-the-art modeling, the good agreement between theory and experiment suggests that the lateral arrangement of the molecule is strong due to the intermolecular interactions; thus the effect of the exact superstructure of the monolayer on the adsorption height should be minimal. This fact has also been indicated in  experimental studies~\cite{Henze:Bauer:etalSS2007,Kilian:Umbach:SokolowskiSS2006}.

We have previously estimated the PBE+vdW$^{\mathrm{surf}}$ adsorption energy of PTCDA on Au(111) to be approximately 2.4 eV per molecule for the case of the adsorbed monolayer~\cite{Ruiz2012} and a value between 2.23 and 2.17 eV for the case of a \emph{single} adsorbed molecule~\cite{Ruiz:Liu:Tkatchenko2016}. In this work, we have calculated the adsorption energy per molecule for two different coverages $\Theta$ of 1.0 and 0.5 ML using the above-mentioned $\bigl(\begin{smallmatrix} 6 & 1 \\ -3 & 5 \end{smallmatrix} \bigr)$ surface unit cell (see Tab.~\ref{Table_PTCDAAu111_2}).
The quantity $E_{\mathrm{ad}}^{\Theta(\mathrm{ML})}$ does not consider the formation of the monolayer in its definition of adsorption energy, whereas  $E_{\mathrm{ad}}^{\Theta(\mathrm{gas})}$ does (see the supplemental material for details on the adsorption model). 

%%%%%%%%%%%%%%%%%
\begin{table}
\begin{center}
\caption{Adsorption energy $E_{\mathrm{ad}}^\Theta$, given in eV, for PTCDA on Au(111) at a coverage $\Theta = 1.0$ ML and $\Theta = 0.5$ ML using the PBE+vdW$^{\mathrm{surf}}$ method. $E^{\mathrm{(ML)}}_{\mathrm{ad}}$ is the adsorption energy calculated with the PTCDA monolayer as reference whereas $E^{\mathrm{(gas)}}_{\mathrm{ad}}$  is the adsorption energy calculated with respect the molecule in gas phase as reference (see the supplemental material for details on the adsorption model).}\label{Table_PTCDAAu111_2}  
\setlength{\tabcolsep}{12pt}
\begin{tabular}{c|cc}
$\Theta$ [ML] & $E_{\mathrm{ad}}^{\Theta(\mathrm{ML})}$  & $E_{\mathrm{ad}}^{\Theta(\mathrm{gas})}$  \\
\hline \noalign{\vskip 1pt} 
1.0   & 2.15 & 3.05   \\
0.5   & 2.27 & 2.50  \\
\hline
\end{tabular}
\end{center}
\end{table}
%%%%%%%%

TPD analysis retrieves the adsorption energy in the limit of low coverage. With this in mind, we have calculated PTCDA on Au(111) with $\Theta$ of 0.60, 0.45, 0.30, and 0.15 ML in order to compare the calculated value of the adsorption energy in the limit of low coverage with the experimental result (see Tab.~\ref{Table_PTCDAAu111_3} and also Ref.~\onlinecite{VRuizPhDThesis}). For these results, we have modelled the system using a larger unit cell with an area of 824~\AA$^2$ and a slab with three Au layers, as described in the supplemental material. Notably, at $\Theta = 0.15$, the difference between $E_{\mathrm{ad}}^{\Theta(\mathrm{ML})}$ and  $E_{\mathrm{ad}}^{\Theta (\mathrm{gas})}$ amounts to only 0.04 eV. We take this coverage value as the limit of low coverage for our calculations. Taking the average value between $E_{\mathrm{ad}}^{\Theta(\mathrm{ML})}$ and $E_{\mathrm{ad}}^{\Theta (\mathrm{gas})}$ at $\Theta = 0.15$, the adsorption energy at the limit of the single molecule is $2.15$~eV.
This value will be slightly increased if we consider a small correction due to the number of layers in the surface slab. In comparison to the experimental result~\cite{Stremlau:PTCDAAu111}, the PBE+vdW$^{\mathrm{surf}}$ adsorption energy is overestimated by approximately 0.20 eV. Our current research indicates that this overestimation is related to the absence of many-body dispersion effects (see Ref.~\onlinecite{Maurer2015}). The inclusion of many-body dispersion effects will reduce the overbinding found in pairwise vdW-inclusive methods, yielding an improvement, for instance, in the adsorption energy. 
%%%%%%
\begin{table}
\centering
\caption{Adsorption energy $E_{\mathrm{ad}}^\Theta$, given in eV, for PTCDA on Au(111) at a coverage $\Theta$ of 1.00, 0.60, 0.45, 0.30, 0.15 ML, and the limit of residual coverage with the PBE+vdW$^{\mathrm{surf}}$ method. Details of the adsorption model can be found in the supplemental material and Ref.~\onlinecite{VRuizPhDThesis}.}\label{Table_PTCDAAu111_3}
\setlength{\tabcolsep}{10pt}
\begin{tabular}{c|cc|c}
 $\Theta \ [\mathrm{ML}]$  & $E_{\mathrm{ad}}^{\Theta(\mathrm{ML})}$  & $E_{\mathrm{ad}}^{\Theta (\mathrm{gas})}$ & Exp.~\cite{Stremlau:PTCDAAu111}  \\
 \hline
        1.00 &           2.07   &    2.97    &--     \\
        0.60 &           2.05   &    2.59    &--      \\
        0.45 &           2.14   &    2.40    &--       \\
        0.30 &           2.16   &    2.26    &--        \\
        0.15 &           2.17   &    2.13    &--         \\
\hline
$\lim {\Theta \to 0}$ & \multicolumn{2}{c|}{2.15}  & $1.93 \pm 0.04$    \\ \hline
\end{tabular}
\end{table}
%%%%%%%%

\subsection{PTCDA on Cu(111)}

\paragraph{Experimental Data}
The experimental information of PTCDA on Cu(111) is not as extensive as in the cases of Ag and Au. The adsorption unit cell of the system is larger in comparison to Ag and Au due to the smaller lattice constant of Cu. It was characterized by \citet{Wagner:Bannani:etalJPCM2007} with STM experiments, where they found two coexisting ordered structures. One corresponds to a $(4 \times 5)$ superstructure and the other one is commensurate with respect to the substrate with two molecules per surface unit cell. The lateral arrangement of the molecules in the monolayer, nevertheless, is not yet fully understood. The bonding distance of the monolayer on Cu(111) was studied by \citet{Gerlach:Sellner:etalPRB2007} using the NIXSW technique. They found the monolayer, in terms of the carbon backbone of the molecule, located at a distance of $2.66 \pm 0.02$~\AA{} with respect to the substrate. Their studies also include the adsorption distances of the chemically inequivalent oxygen atoms in PTCDA. The most striking fact of these studies is that the oxygen atoms are located above the carbon backbone of the molecule. 

\paragraph{Theoretical Data}
We have investigated three possible adsorption structures with the PBE+vdW$^{\mathrm{surf}}$ method, derived by \citet{Romaner:Nabok:etalNJP2009}, which are based on experimental data~\citep{Wagner:Bannani:etalJPCM2007}. These structures correspond to different surface unit cells and lateral placement of the molecules within the monolayer structure. Structure 1 corresponds to a smaller surface unit cell than the one proposed by \citet{Wagner:Bannani:etalJPCM2007}. Structure 2 corresponds to the experimental surface unit cell \cite{Wagner:Bannani:etalJPCM2007}, whereas structure 3 corresponds to a different, plausible surface unit cell with the same area as structure 2. The features of these structures are summarized in the supplemental material.

%%%%%%%
\begin{table}
\begin{center}
\caption[Comparison of PBE+vdW$^{\mathrm{surf}}$ results after relaxation for the three studied structures of PTCDA on Cu(111).]{PTCDA on Cu(111). Experimental results are also shown for comparison~ \cite{Gerlach:Sellner:etalPRB2007}. We use $d_{\rm Th/Exp}$ to denote the averaged vertical adsorption heights of the specific atoms obtained with PBE+vdW$^{\mathrm{surf}}$ calculations and NIXSW studies respectively. The adsorption height is given in~\AA{}.
Adsorption energies are given in eV.
}\label{Table_PTCDACuStruct_1}
\begin{tabular}{c|cccc}
  & \multicolumn{3}{c}{$ d_{\rm Th} $} & $d_{\rm Exp}$~\cite{Gerlach:Sellner:etalPRB2007} \\
   & \multicolumn{3}{c}{Structure}  &   \\
   & 1 & 2 & 3 &  \\
\hline \noalign{\vskip 1pt} 
C total & 2.75 & 2.74 & 2.68 &  $2.66 \pm 0.02$ \\
C$_{\mathrm{peryl}}$ & 2.74 & 2.75 & 2.68 & --\\
C$_{\mathrm{func}}$ & 2.79 & 2.69 & 2.68 & --\\
$\Delta \mathrm{C}$ & $-$0.05 & 0.06 & 0.00 & --               \\
O total & 2.75 & 2.61 & 2.62 & $2.81 \pm 0.03$ \\
O$_{\mathrm{carb}}$ & 2.70 & 2.55 & 2.56 & $2.73 \pm 0.06$ \\
O$_{\mathrm{anhyd}}$ & 2.86 & 2.73 & 2.75 & $2.89 \pm 0.06$ \\
$\Delta \mathrm{O}$ & 0.16 & 0.18 & 0.19 & $0.16 \pm 0.08$ \\
\hline \noalign{\vskip 2pt} 
$E_{\mathrm{ad}}^{\Theta(\mathrm{ML})}$  &  2.55  & 2.48  & 2.78 & -- \\ 
$E_{\mathrm{ad}}^{\Theta(\mathrm{gas})}$ & 3.33 &  3.12  & 3.37 & -- \\
\hline
\end{tabular}
\end{center}
\end{table}

Tab.~\ref{Table_PTCDACuStruct_1} shows the average vertical distance of each species in the PTCDA molecule with respect to the topmost unrelaxed substrate layer. The maximum difference in the adsorption distance of the carbon backbone among the three structures is approximately 0.07~\AA, which is found between structures 1 and 3; while a similar difference of 0.06~\AA{} is found between structures 2 and 3. The calculations show, however, that the adsorption distance in structure 3 agrees remarkably better (deviation of 0.02~\AA) with the NIXSW results. On the other hand, the final positions of the oxygen atoms disagree with the experimental results regardless of the structure of the substrate. The averaged position of the carboxylic oxygen atoms are below the carbon backbone, which is in contrast to the findings of \citet{Gerlach:Sellner:etalPRB2007}.

We have also investigated the stability of each structure by calculating the adsorption energy for each case at monolayer coverage. We have considered both the monolayer and the gas phase molecule as reference. The results are summarized in Tab.~\ref{Table_PTCDACuStruct_1}, showing that structure 3 is the most favorable with respect to the adsorption energy per molecule $E^{\mathrm{(ML)}}_{\mathrm{ad}}$ with the monolayer as reference; while structures 1 and 2 are nearly degenerate. 

The fact that the formation of the monolayer from gas phase brings the adsorption energy closer together in the three cases and the structural differences observed in the structures investigated are evidence that the influence of the lateral placement of the molecules and its relation to the surface unit cell cannot be ignored. A more in-depth research in this regard is part of ongoing efforts. The structural and energetic results also suggest that effects beyond the atomic scale might be at play in the monolayer formation of PTCDA on Cu(111), for example, the statistical average of ordered structures which have subtle structural differences.

\section{Nitrogen-containing Systems on Metal Surfaces}
\label{chapter-nitrogen}
\subsection{Azobenzene on Ag(111) and Au(111) surfaces}

\paragraph{Experimental Data}
Azobenzene (AB, see Fig.~\ref{fig-summary}(i)) and its derivatives adsorbed at metal surfaces have been extensively studied experimentally because of the potential use of its photo-isomerization ability in molecular switching devices~\cite{Comstock2005, Alemani2006, Comstock2007, Choi2006}. STM studies have shown that AB molecules adsorbed at Au(111) preferentially adsorb in stripe patterns with molecules stacked orthogonal to the molecular axis~\cite{Comstock2007, Comstock2005, Kirakosian2005}. These parallel stripes can be transformed to a zig-zag phase using STM bias scanning~\cite{Kirakosian2005}. In doing so, a fully packed zig-zag monolayer at higher coverages can be created. TPD measurements reveal desorption temperatures of about 400 and 440~K for AB adsorbed at low coverage on Ag(111) and Au(111), respectively. On both surfaces the molecule forms multilayers and can be desorbed without fragmentation. Using King's method~\cite{King1975} of TPD analysis this amounts to $1.02 \pm 0.06$ and $1.00 \pm 0.15$~eV adsorption energy per molecule on Ag(111) and Au(111)~\cite{Mercurio2010, Schulze2014}. A detailed analysis of adsorption structure and vertical height of AB on Ag(111) has been performed by Mercurio \emph{et al.} using  NIXSW measurements~\cite{Mercurio2010, Mercurio2013}. Using a Fourier vector analysis in conjunction with DFT~\cite{Mercurio2014} it was found that the vertical adsorption height of the central nitrogen atoms (at 210~K) is $2.97 \pm 0.05$~\AA{} and the central dihedral angle between the nitrogen bridge and the neighboring carbon atoms deviates $-0.7 \pm 2.2^{\circ}$ from a flat arrangement. The outer phenyl rings are tilted by $17.7 \pm 2.4^{\circ}$ with respect to the surface as a result of the tilted, stacked AB arrangement. Although no such thorough analysis exists for AB on Au(111), STM topographs suggest a similar arrangement. 

\begin{figure}
\includegraphics[width=\columnwidth]{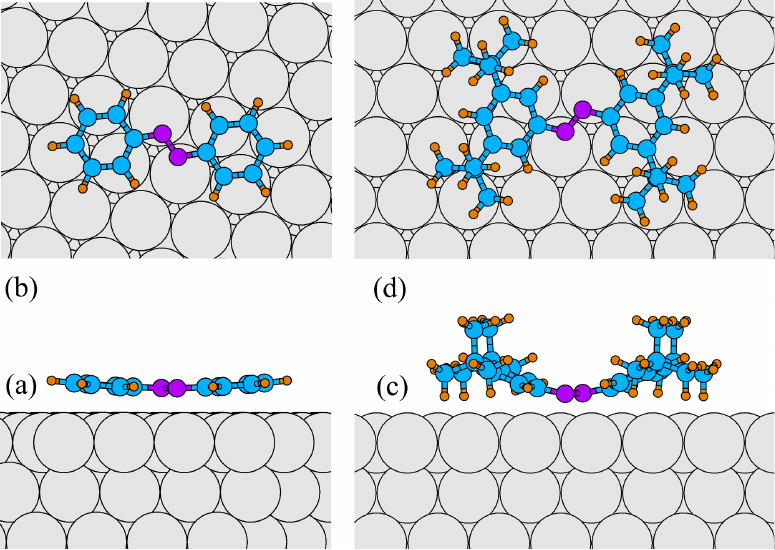}
\caption{\label{fig-azo} Adsorption geometries of AB adsorbed on Ag(111) at the PBE+vdW$^{\mathrm{surf}}$ level in side (a) and top view (b).
TBA adsorbed at Ag(111) in side (c) and top (d) view.}
\end{figure}

\paragraph{Theoretical Data}
AB adsorbed at Ag(111) and Au(111) has been studied in the low coverage limit using non-dispersion corrected PBE~\cite{McNellis2009a}, different dispersion-inclusive functionals incl. PBE+vdW$^{\mathrm{surf}}$~~\cite{McNellis2009,Mercurio2010, Mercurio2013, Mercurio2014, Maurer_Thesis2014}, and vdW-DF~\cite{Li2012}. Adsorption on both surfaces is largely governed by dispersion interactions and description with a pure PBE functional yields almost no interaction energy. Employing pairwise dispersion-correction schemes such as the PBE-D2 scheme~\cite{Grimme2006} leads to a stable adsorption geometry, however at largely overestimated adsorption height and strong overbinding. In the case of AB at Ag(111) described with vdW-DF, the interaction energy is found to be close to experiment (0.98~eV)~\cite{Li2012}, however the vertical adsorption height is 0.5~{\AA} larger than what is found in experiment. The most stable adsorption of AB occurs with the azobridge positioned above a bridge site of the (111)-facet~\cite{McNellis2009a}, the structure is shown in Fig.~\ref{fig-azo}(a) and (b). The adsorption energy and adsorption height as described by PBE+vdW$^{\mathrm{surf}}$ is summarized in Tab.~\ref{tab-azo}. The calculations have been performed in a ($6 \times 4$) surface unit cell using a 4-layer metal slab~\cite{McNellis2009}. For both AB at Ag(111) and Au(111), even with effectively included substrate screening \textit{via} vdW$^{\mathrm{surf}}$, the overbinding is sizeable, as reflected in an underestimation of adsorption height and overestimation of adsorption energies when compared to experiment. The extent of this overestimation is significantly larger for adsorpion on Ag(111) than on Au(111). Mercurio \emph{et al.} have shown that equilibrium geometries at low coverage are not sufficient to model the finite-temperature high-coverage situation in the NIXSW and TPD  measurements of AB on Ag(111)~\cite{Mercurio2013}. When considering higher coverages and correcting for anharmonic changes to the geometry at higher temperature, the agreement between experiment and theory is drastically improved (see Tab.~\ref{tab-azo} line AB/Ag(111) T=210~K). Its molecular flexibility and sizable changes in geometry and interaction energy as a function of coverage make AB an especially challenging benchmark system to an accurate description of dispersion interactions. At the PBE+vdW$^{\mathrm{surf}}$ level, the adsorption energies are overestimated by about 70 to 40\% when compared to TPD experiments. When accounting for finite-temperature effects in the case of AB/Ag(111) the deviations in the description of adsorption energies on both substrates are very similar (33 vs. 40\%). An accurate description of the binding energy as observed in TPD experiments has recently been achieved using explicit ab initio molecular dynamics simulation of the desorption process~\cite{Maurer2016}.

\subsection{TBA on Ag(111) and Au(111) surfaces}
\paragraph{Experimental Data}
3,3',5,5'-tetra-\emph{tert}-butyl-azobenzene (TBA) is an AB derivative substituted with four \emph{tert}-butyl legs (TB legs) (see Fig. \ref{fig-summary} (j) and Fig.~\ref{fig-azo} (c) and (d)) for which successful photo-induced molecular switching has been reported when adsorbed at the Au(111) surface~\cite{Comstock2007, Hagen2007}. STM topographs show distinct island formation suggesting a dominance of lateral interactions over adsorbate-substrate interactions~\cite{Alemani2006, Comstock2007}. The STM topograph of TBA consists of four distinct protrusions representing the TB legs and a depression at the position of the central Azo-bridge. TPD measurements reveal desorption temperatures of about 500 and 540~K corresponding to binding energies of $1.30 \pm 0.20$ and $1.69 \pm 0.15$~eV when adsorbed at Ag(111) and Au(111), respectively~\cite{Schulze2014}. The adsorption geometry of TBA on Ag(111) at room temperature has been determined with NIXSW~\cite{McNellis2010a, Mercurio_PhD_Thesis}. The resulting adsorption height of the central nitrogen atoms is found to be $3.10 \pm 0.06$~\AA{} and the average adsorption height of all carbon atoms is determined as $3.34 \pm 0.15$~\AA{}. From a Fourier vector analysis of the NIXSW signal a flat adsorption geometry with a minimally bent molecular plane is proposed. The extent of signal incoherence suggests that on average 50\% of TB legs are oriented with two methyl groups towards the surface, and 50\% are oriented into the opposite direction~\cite{Mercurio_PhD_Thesis}.

\paragraph{Theoretical Data}
TBA adsorbed at Ag(111) and Au(111) has been modelled in a ($6 \times 5$) surface unit cell with 4 layers of metal~\cite{Maurer_Thesis2014}. The most stable lateral adsorption site, similar as in the AB case is the bridge site, with the molecular axis significantly bent around the central azo-bridge (25 and 14$^\circ$ torsion away from a flat plane on Ag(111) and Au(111)). The TB legs are simulated facing two methyl groups towards the surface. The resulting adsorption energies and geometries are shown in Tab.~\ref{tab-azo}. Whereas the adsorption height (experimentally only measured for TBA/Ag(111)) is in good agreement, adsorption energies are overestimated by about 70 and 30\% for adsorption at Ag(111) and Au(111), which is almost identical to what is found for AB adsorbed at these surfaces. The significantly larger overestimation on Ag(111) suggests, similarly as in the case of AB/Ag(111) an increased relevance of finite-temperature effects leading to an apparent adsorption geometry at larger distances from the surface. The higher relevance of finite-temperature effects on Ag(111) surfaces can be understood by the coexistence of dispersion interactions between the TB legs, the phenyl groups and the substrate and small covalent interaction contributions between the central azo bridge and the metal substrate. 

\begin{table*}
\caption{\label{tab-azo}Adsorption Energy (in eV) and perpendicular heights of Nitrogen atoms 
(in \AA{}) for AB and TBA adsorbated on (111) metal surfaces.
AB/Ag(111) T=210~K refers to simulation data that has been corrected for finite-temperature effects~\cite{Mercurio2013}.}
%\begin{ruledtabular}perpendicular
\begin{tabular}{lcccc}
&\multicolumn{2}{c}{PBE+vdW$^{\rm surf}$} &\multicolumn{2}{c}{Exp.} \\
\cline{2-5} \noalign{\vskip 2pt} 
&\textit{E}$\rm_{ad}$ & \textit{d$_{N-Me}$} & \textit{E}$\rm_{ad}$~\cite{Schulze2014} & \textit{d$_{N-Me}$} \\
\hline  \noalign{\vskip 2pt} 
AB/Ag(111)			&  1.76  & 2.58 & $1.02 \pm 0.06$ & $2.97 \pm 0.05$~\cite{Mercurio2013} \\
AB/Ag(111) T=210~K		&  1.33  & 2.98 & -- & --  \\
AB/Au(111)			&  1.45  & 3.13 & $1.00 \pm 0.15$ & -- \\
TBA/Ag(111)			&  2.28  & 2.36 & $1.30 \pm 0.20$ & $3.10 \pm 0.06$~\cite{McNellis2010a} \\
TBA/Au(111)			&  2.24  & 2.91 & $1.69 \pm 0.15$ & -- \\ \hline
\end{tabular}
\end{table*}
%%%%%%%%%%%%%%%%%%%%%%%%%%%%%%%%%%%%%%%%%%%%%%%%%%

\section{General Discussion}

The adsorption geometry and adsorption energy of an adsorbate on a surface is determined by a number of different contributions: (1) long-range correlations such as dispersion interaction between adsorbate and surface, (2) covalent contributions between chemically active groups at closer distance, (3) the energetic penalty due to geometrical distortion upon adsorption, (4) the Pauli exchange repulsion due to overlapping electron densities, and (5) the attractive or repulsive lateral interactions between adsorbates. All of the above contributions have to be accounted for to arrive at an accurate electronic structure description of HIOSs. The set of systems we have presented in this work covers a large spectrum of interactions, ranging from pure dispersion interactions in the case of rare-gas atoms to highly flexible island-forming heteroaromatic compounds such as azobenzene. In the following we will quantify the reliability of the DFT+vdW$^{\mathrm{surf}}$ method in describing these systems in terms of the mean absolute deviation (MAD) from the above presented experimental reference values (see supplemental material).

Metal surface-adsorbed rare-gas atoms, such as xenon provide the case of pure dispersion interactions of a single atom with a polarizable surface. This is the case for which the vdW$^{\mathrm{surf}}$ method is set out to perform well, which it in fact does throughout the studied surfaces. Adsorption heights are within a MAD of 0.06~\AA{} from experiment, with the closest agreement in the case of Cu(110) and the largest deviation (0.14~\AA) in the case of Cu(111). The same holds for adsorption energies with a MAD of 0.03~eV or 0.8 kcal/mol and the largest error (0.06~eV), again, in the case of Cu(111). This corresponds to a relative error of 12~\% with respect to the magnitude of adsorption energies.

Aromatic adsorbates such as benzene and naphthalene introduce additional aspects such as covalent contributions and geometrical distortions on reactive surfaces such as Pt(111) and a more spatially polarizable electron distribution through the aromatic $\pi$-conjugation of the molecule. Based on the few cases, where experimental references on the adsorption geometry exist (Bz/Ag(111) and Bz/Pt(111)), DFT+vdW$^{\mathrm{surf}}$ appears to correctly describe these systems. The corresponding MAD of 0.06~\AA{} is comparable to the accuracy achieved for xenon adsorption on metals. However, the description of the adsorption energy with DFT+vdW$^{\mathrm{surf}}$ is significantly worse (MAD of 0.23~eV). Throughout all systems DFT+vdW$^{\mathrm{surf}}$ leads to overbinding of the adsorbate, which is more pronounced for the larger Np than for Bz. For the latter the MAD is only 0.09~eV. This may be an indication that the complex long-range interaction of the polarizable adsorbate/substrate complex cannot be mapped onto effective pairwise-additive contributions. 

The overbinding as observed for small and intermediate aromatic molecules seems to be somewhat decoupled from the description of the adsorption geometry. This becomes more evident for the adsorption of extended and compacted nanostructures such as DIP and C$_{60}$. Unfortunately, no experimentally measured adsorption energies exist, however the accuracy of DFT+vdW$^{\mathrm{surf}}$ in describing the adsorption height of these systems is equivalent to the above systems and is reflected in an MAD in adsorption height of 0.05~\AA{}.

Heteroaromatic adsorbates with functional groups that feature lone pairs and unsaturated bonds between carbon and oxygen, sulfur, or nitrogen introduce additional complexity. Upon adsorption these heteroatoms will bond differently to the surface than the carbon backbone. Functional groups with sulfur and oxygen for example may engage in surface bonds that are stronger than the average C--metal interaction. On the other hand nitrogen atoms may exhibit an overall weaker interaction. 

%Thp
For thiophene--metal adsorption we find the adsorbate structure to be dominated by the strong S--metal bond which has a shorter distance to the surface than the ideal vertical adsorption height of the aromatic backbone of the molecule. The result is a tilted, flexible adsorption geometry that may vary considerably as a function of coverage and temperature. For the low-coverage cases with little tilt angle that we reviewed here, DFT+vdW$^{\mathrm{surf}}$, performs similarly as for benzene and naphthalene. The deviation in adsorption height for Thp on Cu(111) is 0.16~\AA{} and the MAD in adsorption energy for the three coinage metal substrates is 0.14~eV when compared to experimental adsorption energies extracted from TPD \textit{via} the Redhead technique. At this point it should be mentioned that this method of calculating the adsorption energy from the desorption temperature and a range of assumed pre-exponential factors is highly disputed and only used due to lack of more recent measurements. More accurate experimental reference values are in dire need.

%PTCDA
In the case of PTCDA, a molecule with a conjugated backbone and terminal di-oxo anhydride groups, adsorbed on different facets of coinage metal surfaces, we find two strongly different spatially separated chemical moieties. The terminal oxo groups engage with the surface in a covalent bond, whereas the interaction of backbone and surface is mainly dominated by dispersion interactions. The active terminal groups in addition lead to lateral interactions that enable a multitude of different stable overlayer phases. However, even for this rather complicated case DFT+vdW$^{\mathrm{surf}}$ yields an MAD in adsorption height of 0.08~\AA{} that is comparable to the above systems featuring considerably simpler chemistry. The largest deviation can be found for the O--Cu distance in PTCDA/Cu(111) with a discrepancy of 0.19~\AA{} when compared to experiment. An experimental estimate for the adsorption energy only exists for PTCDA/Au(111), which is overestimated by DFT+vdW$^{\mathrm{surf}}$ by 0.22~eV. The adsorption energy of PTCDA/Ag(111) has recently been estimated from TPD measurements of a smaller analogue called NTCDA~\cite{Maurer2015} to be in the range of 1.4 to 2.1~eV. The DFT+vdW$^{\mathrm{surf}}$ value for dilute coverage is larger than this estimate. However, there are several different disputed estimates in literatute~\cite{Maurer2015,Ruiz2012,Berland2014a}. An accurate experimental reference value for the adsorption energy of PTCDA on Ag(111) that can settle this dispute remains to be measured.

%PTCDA
Another important aspect of functional groups on the adsorption of aromatic molecules is the emergence of lateral intermolecular interactions such as hydrogen bonds. In the case of PTCDA, the carboxylic oxygens in the molecule form hydrogen bonds with the neighboring molecules and thereby generate the herringbone pattern that is characteristic of the monolayer formation of PTCDA on coinage metals. This fact is initially independent of the interaction with the substrate since PTCDA forms crystals with layered molecular stacks in which the ordering pattern of the molecules in each layer is that of a herringbone arrangement closely related to the one found in the monolayer. There is, nevertheless, an effect on the ordering of the monolayer on each substrate which depends on the adsorbate-substrate interaction. This can be related to the degree of commensurability between the lattice parameters of the substrate bulk and those of the stacking layer of the organic crystal. This leads to the formation of a commensurate monolayer on Ag(111) and a point-on-line coincidence on Au(111). In the case of Cu(111), the mismatch leads to different adsorption sites for oxygen atoms yielding a different distortion in each molecule forming the monolayer, which is reflected in the large discrepancy between experiment and calculation and the low coherent fraction for the carboxylic oxygens found in the NIXSW experiments~\cite{Gerlach:Sellner:etalPRB2007}. Metal-adsorbed PTCDA also nicely exemplifies the importance of intermolecular interactions and their contribution to the total adsorption energy. As shown above the adsorption energy varies strongly as a function of coverage and a correct simulation model of the experimental overlayer structure is extremely important to arrive at an accurate first-principles adsorption energy.

%Azo
Contrary to the cases of Thp and PTCDA, the diazo groups in AB and TBA do not strongly interact with Ag(111) and Au(111) surfaces and also contribute to the $\pi$-conjugation of the molecule. More importantly, the central azo bridge in the molecule makes it inherently flexible, which translates into a strongly anharmonic binding component in the adsorbate/substrate complex~\cite{Mercurio2013}. The result is a strong temperature dependence of adsorbate geometry and energetics and a large discrepancy in both height and adsorption energy. MADs for AB and TBA 0~K adsorption models are 0.57~\AA{} and 0.68~eV, respectively, when compared to finite-temperature experimental results. Whereas anharmonic finite-temperature correction of the adsorbate geometry~\cite{Mercurio2013} leads to an excellent agreement of simulated and measured adsorption height, the adsorption energy remains 0.31~eV overestimated -- a deviation that is in line with the above studied systems.

When we combine the DFT+vdW$^{\mathrm{surf}}$ results for the whole dataset of vertical adsorption heights and adsorption energies we arrive at MADs of 0.11~\AA{} and 0.26~eV. However, if we exclude the 0~K models of metal-adsorbed AB and TBA and only include the finite-temperature corrected AB/Ag(111) the MADs are 0.06~\AA{} for 23 different vertical adsorption heights and 0.16~eV for 17 different adsorption energies (see supplemental material for details). The latter results reflect the same accuracy throughout the dataset and support the assessment that DFT+vdW$^{\mathrm{surf}}$ yields a reliable description of adsorption geometries, however, at the same time appears to systematically overestimate adsorption energetics of systems more complex than rare-gas atoms. The here presented dataset of 23 vertical adsorption heights and 17 adsorption energies of different HIOSs establishes a comprehensive benchmark and may serve as a tool for the assessment of current and future electronic structure methods.

%%%%%%%%%%%%%%%%%%%%%%%%%%%%%%%%%%%%%%%%%%%%%%%%%%

\section{Conclusions and Outlook}
We have presented results for the structure and stability of a series of HIOSs using a dispersion-inclusive DFT-based method that can reliably describe a wide range of interactions including covalent  bonding, electrostatic interactions, Pauli repulsion, and vdW interactions. The noticeable improvement in the calculation of adsorption distances and energies that we have observed with the DFT+vdW$^{\mathrm{surf}}$ method compared to other pairwise-additive dispersion-corrected methods indicates the importance of the inclusion of the collective screening effects present in the substrate for the calculation of vdW interactions, with particular importance in the case of organic-metal interfaces. From a general perspective, however, there are still many important aspects left to consider in order to achieve both quantitative and predictive power in the simulation of the structure and stability of complex interfaces.

Throughout the here discussed dataset, DFT+vdW$^{\mathrm{surf}}$ yields a slight overestimation of adsorption energies. We relate this to the fact that the complex adsorbate/substrate interactions in systems beyond simple rare-gas atoms cannot be effectively captured in a pairwise-additive dispersion scheme and a more explicit account of many-body long-range correlation contributions is necessary.
This could be achieved by incorporating the collective response of the combined adsorbate/substrate system, rather than the effective inclusion of the substrate response alone.~\citep{Liu:Tkatchenko:SchefflerAccChemRes2014} High-level quantum-chemistry methods or many-body methods such as the Random Phase Approximation (RPA) for the correlation energy could be used for this purpose. These approaches, however, either perform well for isolated molecules or periodic surfaces, but are rarely applicable for the combined system. Recent results on Xe and PTCDA adsorption at Ag(111)~\cite{Maurer2015} suggest that the many-body description of dispersion interactions at the level of the DFT+MBD method~\cite{Tkatchenko2012,DiStasio2014} may be an efficient approach to remedy the intrinsic overbinding in DFT+vdW$^{\mathrm{surf}}$. The DFT+MBD method may thereby significantly improve the description of adsorption energies, geometries of flexible systems, and vibrational properties of HIOSs.

Another equally important challenge for future methods will be a more sophisticated account of local correlation and exchange effects beyond a semi-local exchange-correlation treatment. Range-separated hybrid functionals~\cite{Atalla2013}, screened second order exchange methods~\cite{Paier2006}, and recent many-body perturbation theory approaches~\cite{Ren2011,Caruso2013,Ren2013} in combination with methods that provide long-range dispersion interactions may provide the necessary accuracy to correctly reproduce the charge-distribution at the molecule/surface interface, the molecular level alignment, and the correct surface potential decay.

The here reviewed representative set of 23 adsorption heights and 17 adsorption energies for HIOSs may provide a useful tool in evaluation and assessment of methodological improvements and may also be extended to measurable electronic, vibrational, and spectroscopic properties. For this dataset, the PBE+vdW$^{\mathrm{surf}}$ method yields a MAD in adsorption height of 0.06~\AA{} and a MAD in adsorption energy of 0.16~eV (3.7~kcal/mol) and represents the current state-of-the-art in electronic structure description of HIOSs.
%%%%%%%%%%%%%%%%%%%%%%%%%%%%%%%%%%%%%%%%%%%%%%%%%%

\section*{Acknowledgements}
Support from the DFG project RE1509/16-02, the DFG-SFB/951 HIOS project A10, the European Research Council (ERC-StG VDW-CMAT), and the DoE - Basic Energy Sciences grant no. DE-FG02-05ER15677 is acknowledged for this work. The authors thank Petra Tegeder for helpful comments on section~\ref{chapter-experiment}.
%%%%%%%%%%%%%%%%%%%%%%%%%%%%%%%%%%%%%%%%%%%%%%%%%%
%%%%%%%%%%%%%%%%%%%%%%%%%%%%

\section*{References}
\bibliographystyle{apsrev4-1}
% \bibliography{references}

\providecommand{\noopsort}[1]{}\providecommand{\singleletter}[1]{#1}%

\end{document}